\documentclass[a4paper]{article}
\usepackage[aux]{rerunfilecheck}
\usepackage[utf8]{inputenc}
\usepackage[T1]{fontenc}
\usepackage[english]{babel}
\usepackage[margin=2.5cm]{geometry}
\usepackage{xcolor}
\usepackage{csquotes}
\usepackage{amsmath}
\usepackage{textcomp}
\usepackage{amssymb}
\usepackage{amsfonts}
\usepackage{hepnicenames}
\usepackage{hepparticles}
\usepackage{mathtools}
\usepackage{microtype}
\usepackage{listings}
\usepackage{multirow}
\usepackage{graphicx}
\usepackage{grffile}
\usepackage{longtable}
\usepackage{wrapfig}
\usepackage{rotating}
\usepackage[normalem]{ulem}
\usepackage{capt-of}
\usepackage[
round-mode=places,
round-precision=1,
binary-units=true,
]{siunitx}
\usepackage{booktabs}
\usepackage[colorlinks=true]{hyperref}
\usepackage{cleveref}
\usepackage[backend=biber,sorting=none]{biblatex}
\usepackage{float}
\usepackage[section, below]{placeins}
\usepackage{caption}

\makeatletter
\newif\if@preliminary
\@preliminaryfalse
\def\preliminary{\@preliminarytrue}
\def\preprintno#1{\def\@preprintno{#1}}
\def\address#1{\def\@address{#1}}
\def\email#1#2{\thanks{\tt #1@{}#2}}
\def\abstract#1{\def\@abstract{#1}}
\renewcommand\abstractname{ABSTRACT}
\newlength\preprintnoskip
\newlength\abstractwidth
\@titlepagetrue
\renewcommand\maketitle{\begin{titlepage}%
  \let\footnotesize\small
  \hfill\parbox{\preprintnoskip}{%
  \begin{flushright}\@preprintno\end{flushright}}\hspace*{1cm}
  \vskip 60\p@
  \begin{center}%
    {\Large\bf\boldmath \@title \par}\vskip 1cm%
    {\sc\@author \par}\vskip 3mm%
    {\@address \par}%
    \if@preliminary
      \vskip 2cm {\large\sf PRELIMINARY DRAFT \par \@date}%
    \fi
  \end{center}\par
  \@thanks
  \vfill
  \begin{center}%
    \parbox{\abstractwidth}{\centerline{\abstractname}%
    \vskip 3mm%
    \@abstract}
  \end{center}
  \end{titlepage}%
  \setcounter{footnote}{0}%
  \let\thanks\relax\let\maketitle\relax
  \gdef\@thanks{}\gdef\@author{}\gdef\@address{}%
  \gdef\@title{}\gdef\@abstract{}\gdef\@preprintno{}
}%
\def\@citex[#1]#2{\if@filesw\immediate\write\@auxout{\string\citation{#2}}\fi
  \def\@citea{}\@cite{\@for\@citeb:=#2\do
    {\@citea\def\@citea{,\penalty\@m}\@ifundefined
       {b@\@citeb}{{\bf ?}\@warning
       {Citation `\@citeb' on page \thepage \space undefined}}%
\hbox{\csname b@\@citeb\endcsname}}}{#1}}
\def\citerange{\@ifnextchar [{\@tempswatrue\@citexr}{\@tempswafalse\@citexr[]}}
\def\@citexr[#1]#2{\if@filesw\immediate\write\@auxout{\string\citation{#2}}\fi
  \def\@citea{}\@cite{\@for\@citeb:=#2\do
    {\@citea\def\@citea{--\penalty\@m}\@ifundefined
       {b@\@citeb}{{\bf ?}\@warning
       {Citation `\@citeb' on page \thepage \space undefined}}%
\hbox{\csname b@\@citeb\endcsname}}}{#1}}
\long\def\@makecaption#1#2{%
  \vskip\abovecaptionskip
  \sbox\@tempboxa{#1: \emph{#2}}%
  \ifdim \wd\@tempboxa >\hsize
    #1: \emph{#2}\par
  \else
    \hbox to\hsize{\hfil\box\@tempboxa\hfil}%
  \fi
  \vskip\belowcaptionskip}
\makeatother

\newcommand\CO{{\cal O}}

\newcommand\be{\begin{equation}}
\newcommand\ee{\end{equation}}
\newcommand\bea{\begin{eqnarray}}
\newcommand\eea{\end{eqnarray}}
\newcommand\ba{\begin{array}}
\newcommand\ea{\end{array}}

\newcommand{\dd}{\mathrm{d}}

\newcommand\sqrts{\ensuremath{\sqrt{s}}\xspace}

\newcommand{\wz}{\textsc{Whizard}\xspace}

\newcommand{\sindarin}{\textsc{Sindarin}\xspace}
\newcommand{\vegas}{\textsc{Vegas}\xspace}
\newcommand{\vamp}{\textsc{Vamp}\xspace}

\newcommand{\om}{\textsc{O'Mega}\xspace}
\newcommand{\ol}{\textsc{OpenLoops}\xspace}

\newcommand{\re}{\textsc{RECOLA}\xspace}

\newcommand{\go}{\textsc{GoSam}\xspace}

\newcommand{\pythia}{\textsc{Pythia}\xspace}

\newcommand{\feynrules}{\textsc{FeynRules}\xspace}

\newcommand{\Rset}{\mathbb{R}}

\newcommand{\fortran}{\texttt{Fortran}}
\newcommand{\ffortran}{\texttt{Fortran2008}}
\graphicspath{{../},{.},{plots}}
\DeclareGraphicsExtensions{{.pdf}}

\newcommand\Pjet{\HepGenParticle{j}{}{}}

\begin{document}

\preprintno{%
  DESY 18-089 \\
  SI-HEP-2018-32
}

\title{Parallel Adaptive Monte Carlo Integration with the Event Generator WHIZARD}

\author{
  Simon Brass\email{simon.brass}{uni-siegen.de}$^{a}$,
  Wolfgang Kilian\email{kilian}{physik.uni-siegen.de}$^{a,d}$,
  J\"urgen Reuter\email{juergen.reuter}{desy.de}$^b$,
}  

\address{\it%
    $^a$University of Siegen, Department of Physics,
    D--57068 Siegen, Germany, \\
    $^b$DESY Theory Group,
    D--22607 Hamburg, Germany
}

\date{%
  \today%
  }

\abstract{
  We describe a new parallel approach to the evaluation of phase space
  for Monte-Carlo event generation, implemented within the framework
  of the \wz\ package.  The program realizes a twofold self-adaptive
  multi-channel parameterization of phase space and makes use of the
  standard \texttt{OpenMP} and \texttt{MPI} protocols for
  parallelization.   The modern \texttt{MPI3} feature of asynchronous
  communication is an essential ingredient of the computing model.
  Parallel numerical evaluation applies both to phase-space integration
  and to event generation, thus covering the most computing-intensive
  parts of physics simulation for a realistic collider environment.
}
\maketitle
%%\clearpage

\setcounter{tocdepth}{2}
\tableofcontents

%%%%%%%%%%%%%%%%%%%%%%%%%%%%%%%%%%%%%%%%%%%%%%%%%%%%%%%%%%%%%%%%%%%%%%%%
\section{Introduction}
\label{sec-1}

Monte-Carlo event generators are an indispensable tool
of elementary particle physics. Comparing collider data with
a theoretical model is greatly facilitated if there is a sample of
simulated events that represents the theoretical prediction, and can
directly be compared to the event sample from the particle detector.  The
simulation requires two steps: the generation of particle-level
events, resulting in a set of particle species and momentum
four-vectors, and the simulation of detector response.  To generate
particle-level events, a generator computes partonic observables and
partonic event samples which then are dressed by parton shower,
hadronization, and hadronic decays.  In this paper, we focus on the
efficient computation of partonic observables and events. 

Hard scattering processes involve
Standard-Model (SM) elementary particles -- quarks, gluons, 
leptons, $\PWpm$, $\PZ$, and Higgs bosons, and photons.  The
large number and complexity of scattering events recorded at detectors
such as ATLAS or CMS, call for a matching computing power in
simulation. Parallel evaluation that makes maximum use of available
resources is an obvious solution.

The dominant elementary processes at the Large Hadron Collider (LHC)
can be described as $2\to 2$ or $2\to 3$ particle production, where
resonances in the final state subsequently decay, and additional jets
can be accounted for by the parton shower.  Cross sections and
phase-space distributions are available as analytic expressions. Since
distinct events are physically independent of each other, parallel
evaluation is done trivially by generating independent event samples
on separate processors.  In such a situation, a parallel simulation
run on a multi-core or multi-processor system can operate close to
optimal efficiency.

However, the LHC does probe
rarer partonic processes which are of the type $2\to n$ where $n\geq
4$.  There are increasing demands on the precision in data analysis at the
LHC and, furthermore, at the planned future high-energy and
high-luminosity lepton 
and hadron colliders.  This forces the simulation to
go beyond the leading order in perturbation theory, beyond the
separation of production and decay, and beyond simple approximations
for radiation.  For instance, in processes such as top-quark
pair production or vector-boson scattering,
the simulation must
handle elementary $n=6,8,10$ processes.  Closed analytical expressions for
arbitrary phase-space distributions are not available.

Event generators for particle physics processes therefore rely on
statistical methods both for event generation and for the
preceding time-consuming numerical integration of the phase space.
All major codes involve a Monte-Carlo rejection algorithm for
generating unweighted event samples.  This requires  knowledge of the
total cross section and of a reference limiting distribution over the
whole phase space.  Calculating those quantities relies again on
Monte-Carlo methods, which typically involve an adaptive iteration
algorithm.  A large fraction of the total computing time
cannot be trivially parallelized since it involves significant
communication.  Nevertheless, some of the main MC event generators
have implemented certain parallelization features, e.g.,
\texttt{Sherpa}~\cite{Gleisberg:2008ta} and 
\texttt{MG5\_aMC@NLO}~\cite{Alwall:2014hca}.

In the current paper, we describe a new approach to efficient
parallelization for adaptive Monte Carlo integration and
event generation, implemented within the \wz~\cite{Kilian:2007gr}
Monte-Carlo integration 
and event-generation program. The approach combines independent
evaluation on separate processing units with asynchronous
communication via MPI 3.0 with internally parallelized loops
distributed on multiple cores via OpenMP.  In Sec.~\ref{sec:whizard},
we give an overview of the workflow of the \wz\ event generation
framework with the computing-intensive tasks that it has to perform.
Sec.~\ref{sec:vegas} describes the actual MC integration and event
generation algorithm.  The parallelization methods and necessary
modifications to the algorithm are detailed in
Sec.~\ref{sec:parallel}.  This section also shows our study on the
achievable gain in efficiency for typical applications in high-energy
physics. Finally, we conclude in Sec.~\ref{sec:conclusions}.

%%%%%

\section{The WHIZARD multi-purpose event generator framework}
\label{sec:whizard}

We will demonstrate our parallelization algorithms within the
\wz\ framework~\cite{Kilian:2007gr}. \wz\ is a multi-purpose Monte-Carlo
integration and event generator program.  In this Section, we describe
the computing-intensive algorithms and tasks which are potential targets
of improvement via parallel evaluation. In order to make the section
self-contained, we also give an overview of the capabilities of \wz.

The program covers the complete workflow of particle-physics
calculations, from setting up a model Lagrangian to generating
unweighted hadronic event samples.  To this end, it combines internal
algorithms with external packages.  Physics models are available either as
internal model descriptions or via interfaces to external packages,
e.g. for \feynrules~\cite{Christensen:2010wz}.  For any model,
scattering amplitudes are automatically constructed and accessed for
numerical evaluation via the included matrix-element
generator
\om~\cite{Moretti:2001zz,Ohl:2002jp,Ohl:2004tn,Hagiwara:2005wg,Kilian:2012pz,Chokoufe_Nejad_2015,}.
The calculation of partonic cross sections, observables, and events is
handled within the program itself, as detailed below.  Generated
events are showered by internal routines~\cite{Kilian:2011ka}, showered and
hadronized by standard interfaces to external programs, or by means of
a dedicated interface to \pythia~\cite{Sjostrand:2006za,Sjostrand:2014zea}.  
For next-to-leading-order (NLO) calculations, \wz\ takes virtual
and color-/charge-/spin-correlated matrix elements from the
one-loop providers \ol~\cite{Cascioli:2011va,Buccioni:2017yxi},
\go~\cite{Cullen:2011ac,Cullen:2014yla}, or \re~\cite{Actis:2016mpe},
and handles subtraction within the Frixione-Kunszt-Signer scheme
(FKS)~\cite{Frixione:1995ms,Frederix_2009,Jezo:2015aia,Reuter_2016}.
Selected \wz\ results at NLO, some of them obtained using parallel
evaluation as presented in this paper, can be found in
Refs.~\cite{Kilian:2006cj,Binoth:2009rv,Greiner:2011mp,Nejad:2016bci, 
Bach:2017ggt}. 

The core part of \wz\ is the phase-space integration and its
parameterization in terms of importance-ordered channels. The phase
space is the manifold of kinematically allowed energy-momentum
configurations of the external particles in an elementary process.
User-defined cuts may additionally constrain phase space in
rather arbitrary ways. \wz\ specifically allows for arbitrary
phase-space cuts, that can be steered from the input file via its
scripting language \sindarin\ without any (re-)compilation of
code.  The program determines a set of phase-space parameterizations
(called \emph{channels}), i.e., bijective mappings of a 
subset of the unit $d$-dimensional hypercube onto the phase-space
manifold.  For the processes of interest, $d$ lies between 2 and some
25 dimensions. Note that the parameterization of non-trivial beam structure
in form of parton distribution functions, beam
spectra, electron structure functions for initial-state radiation,
effective photon approximation, etc., provides additional
dimensions to the numerical integration.  The actual integrand, i.e.,
the square of a 
transition matrix element evaluated at the required order in
perturbation theory, is defined as a function on this cut phase space.
It typically 
contains sharp peaks (resonances) and develops
poles in the integration variables just beyond the boundaries.  We 
collectively denoted those as "singularities" in a slight abuse of
language.  In effect, the numerical value of the integrand varies over
many orders of magnitude. 

For an efficient integration, it is essential that the program
generates multiple phase-space channels for the same process.  Each
channel has the property that a particular subset of the singularities
of the integrand maps to a slowly varying distribution (in the ideal
case a constant) along a coordinate axis of the phase-space manifold
with that specific mapping.  The set of channels has to be large
enough that it  covers all numerically relevant singularities.  This
is not a well-defined requirement, and \wz\ contains a heuristic
algorithm that determines this set.  The number of channels may range
from one (e.g. $\Pep\Pem \to \APmuon\Pmuon$ at $\sqrts = 40$ GeV, no beam
structure) to some $10^6$ (e.g. vector boson scattering at the LHC, or
BSM processes at the LHC) for typical applications.

The actual integration is done by the \vamp\ subprogram, which is a
multi-channel version of the self-adaptive \vegas\ algorithm.  We
describe the details of this algorithm below in Sec.~\ref{sec:vegas}.
In essence, for each channel, the hypercube is binned along each
integration dimension, and the bin widths as well as the channel
weight factors are then iteratively adapted in order to reduce the
variance of the integrand as far as possible.  This adaptive
integration step includes the most computing-intensive parts of \wz.
CPU time is spent mostly in (i) the evaluation of the matrix element
at each phase space point which becomes particularly time-consuming
for high-multiplicity or NLO processes, (ii)
the evaluation of the phase-space mapping and its inverse for each
channel, alongside with the Jacobian of this mapping, (iii) sampling
the phase-space points and collecting the evaluation results, and (iv)
the adaption of the bin widths and channel weights.  If the adaptation
is successful, it will improve the integration result: the relative
error of the Monte-Carlo integration is estimated as
\begin{equation}\label{eq:error-scaling}
  \frac{\Delta I}{I} = \frac{a}{\sqrt{N}},
\end{equation}
where $I$ is the integral that has to be computed, $\Delta I$ is the
statistical error of the integral estimate, and $N$ is the number of
phase-space points for which the integrand has been evaluated.
Successful adaptation will improve the integral error by reducing the
\emph{accuracy} parameter $a$, often by several orders of magnitude.

The program records the integration result.  The accompanying set of
adapted bin widths, for each dimension and for each channel, together
with the maximum value of the mapped integrand, is called an
integration \emph{grid}, and is also recorded.  The set of adapted
grids can then be used for the final simulation step.  Simulation
implies generating a sample of statistically independent events
(phase-space points), with a probability distribution as given by the
integrand value.  After multiple adaptive iterations, the effective
event weights should vary much less in magnitude compared to the
original integrand defined on the phase-space manifold.  Assuming that
the maximum value can be estimated sufficiently well, a simple
rejection algorithm can convert this into a set of unweighted events.
An unweighted event sample constitutes an actual simulation of an
experimental outcome.  The \emph{efficiency} $\epsilon$ of this
reweighting is the ratio of the average event weight over the maximum
weight,
\begin{equation}\label{eq:efficiency}
  \epsilon = \frac{\langle w\rangle_N}{w_{\text{max}}}.
\end{equation}
Since the integrand has to be completely evaluated for each event
before acception or rejection, the unweighting efficiency $\epsilon$
translates directly into CPU time for generating an unweighted event
sample.  Successful adaptation should increase $\epsilon$ as far as
possible.  We note that event generation, for each phase-space point,
involves all channel mappings and Jacobian calculations in the same
way as integration does.

Finally, for completeness, we note that \wz\ contains additional
modules that implement other relevant physical effects, e.g.,
incoming-beam structure, polarization, factorizing processes into
production and decay, and modules that prepare events for actual
physics studies and analyses.  To convert partonic events into
hadronic events, the program provides its own algorithms together
with, or as an alternative to, external programs such as \pythia.
Data visualization and analysis can be performed by its own routines
or by externally operating on event samples, available in various
formats.

Before we discuss the parallelization of the phase-space integration,
in the next section, Sec.~\ref{sec:vegas}, we explain in detail how
the MC integration of \wz\ works.

%%%%%

\section{The MC integrator of \wz: the \vamp\ algorithm}
\label{sec:vegas}

The implementation of the integration and event generation modules of 
\wz\ is based on the \vegas\ algorithm~\cite{Peter_Lepage_1978,
Lepage:123074}.  \wz\ combines the \vegas\ method of adaptive
Monte-Carlo integration with the principle of multi-channel
integration~\cite{Kleiss_1994}.  The basic algorithm and a sample 
implementation have been published as \vamp\ (Vegas AMPlified)
in~\cite{Ohl_1999}.  In this section, in order to make the discussion
of our parallelized re-implementation of the \vamp\ algorithm, we
discuss in detail the algorithm and its application to phase-space
sampling within \wz. Our parallelized implementation for the purpose
of efficient parallel evaluation is then presented in
Sec.~\ref{sec:parallel}.

\subsection{Integration by Monte-Carlo Sampling}

We want to compute the integral $I$ for an integrand $f$ defined on a
compact $d$-dimensional phase-space manifold $\Omega$.  The integrand
$f$ represents a real-valued squared matrix element (in NLO
calculations, this can be a generalization that need not be positive
semidefinite) with potentially high numerical variance.  The
coordinates $p$ represent four-momenta  and, optionally, extra
integration variables such as structure function parameters like
parton energy fractions. The phase-space manifold and the measure
$\dd\mu(p)$ are determined by four-momentum conservation, on-shell
conditions, and optionally by user-defined cuts and weight factors:
\begin{equation}
  I_\Omega[f] = \int_\Omega \dd\mu(p)\,f(p) \, .
\end{equation}
A phase-space parameterization is a bijective mapping $\phi$ from a
subset $U$ of the $d$-dimensional unit hypercube, $U\subset (0,1)^d$,
onto $\Omega=\phi(U)$ with Jacobian $\phi'=\det(\dd\phi/\dd x)$,
\begin{equation}
  p = \phi(x),
  \qquad
  \dd\mu(p) = \phi'(x)\,\dd\mu(x) = \phi'(x)\,\rho_\phi(x)\,\dd^d x \, .
\end{equation}
The measure in this chart becomes proportional to the canonical
measure on $\Rset^d$ with density $\phi'(x)\,\rho_\phi(x)$.  Mappings
$\phi$ may be chosen a priori such that the density
expressed in these coordinates does not exhibit high numerical
variance; for instance, the Jacobian $\phi'$ may cancel integrable
singularities associated with Gram determinants near the edges of
phase space.  Furthermore, we may extend the $x$ integration over the
complete unit hypercube, continuing $\Omega$ arbitrarily while setting
$\dd\mu(x)=0$ for values $x$ outside the integration domain~$U$
(e.g. outside a fiducial phase-space volume).

The basic idea of Monte-Carlo integration~\cite{james1980monte} builds
upon the observation that the integral $I_\Omega[f]$ can be estimated
by a finite sum, where the estimate is given by
\begin{equation}
  E_N[f]
  =
  \langle f \rangle_N
  =
  \frac{1}{N}\sum_{i=1}^N f(\phi(x_i))\,\phi'(x_i)\,\rho_\phi(x_i),
\end{equation}
if the points $x_i$ are distributed according to a uniform random
distribution within the hypercube.  $N$ is the number of random number
configurations for which the integrand has been evaluated, the
\emph{calls}. The integration method is also known as \emph{importance
sampling}.

Asymptotically, the estimators $E_N[f]$ for independent random
sequences $\{x_i\}$ will themselves be statistically distributed
according to a Gaussian around the mean value $I_\Omega[f]$.  The
statistical error of the estimate can be given as the square root of
the variance estimate,
\begin{equation}
  V_N[f]
  =
  \frac{N}{N-1}
  \left(\langle f^2 \rangle_N - \langle f \rangle_N^2\right),
\end{equation}
which is calculated alongside with the integral estimate.
Asymptotically, the statistical error of the integration scales
according to Eq.~\eqref{eq:error-scaling}, where the accuracy $a$
depends on the actual variance of the effective integrand
\begin{equation}
  f_\phi(x)
  =
  f(\phi(x))\,\phi'(x)\,\rho_\phi(x)
\end{equation}
In short, the computation consists of a sequence of \emph{events},
points $x_i$ with associated weights $w_i=f_\phi(x_i)$.

This algorithm has become a standard choice for particle-physics
computations, because (i) the error scaling law $\propto 1/\sqrt{N}$
turns out to be superior to any other useful algorithm for large
dimensionality of the integral $d$; (ii) by projecting on observables
$\CO(\phi(x))$, any integrated or differential observable can be
evaluated from the same event sample; and (iii) an event sample can be
unweighted to accurately simulate the event sample resulting from an
actual experiment. The unweighting efficiency $\epsilon$ as in
Eq.~\eqref{eq:efficiency} again depends on the behavior of the 
effective integrand. 

The optimal values $a=0$ and $\epsilon=1$ are reached if
$f(\phi(x))\,\phi'(x)\,\rho_\phi(x)\equiv 1$.  In one dimension, this
is possible by adjusting the mapping $\phi(x)$ accordingly.  The
Jacobian $\phi'$ should cancel the variance of the integrand $f$, and
thus will assume the shape of this function.  In more than one
dimension, such mappings are not available in closed form, in general.

In calculations in particle-physics perturbation theory,
the integrand $f$ is most efficiently derived recursively, e.g. from
on-particle off-shell wavefunctions like in~\cite{Moretti:2001zz}. The
poles in these recursive structures are the resonant Feynman
propagators. In particular, if for a simple process only a single
propagator contributes, there are standard mappings $\phi$ such that
the mapped integrand factorizes into one-dimensional functions, and
the dominant singularities are canceled.  For this reason, the phase
space channels of \wz\ are constructed from dominating proagators,
which are represented by typical Feynman graphs (even if the matrix
elements do not rely on the redundant expansion into Feynman graphs).
If several graphs contribute, mappings that cancel the singularities
are available only for very specific cases such as massless QCD
radiation, e.g.~\cite{Draggiotis:2000gm,vanHameren:2000aj}. In any
case, we have to deal with some remainder variance that is not
accounted for by standard mappings, such as polynomial factors in the
numerator, higher-order contributions, or user-defined cuts which do
not depend on the integrand~$f$.

\subsection{The \vegas\ algorithm: importance sampling}

The \vegas\ algorithm~\cite{Lepage:123074} addresses the frequent
situation that the effective integrand $f_\phi$ is not far from
factorizable form, but the capabilities of finding an optimal mapping
$\phi$ in closed form have been exhausted.  In that case, it is
possible to construct a factorizable
\emph{step} mapping that improves accuracy and efficiency beyond
the chosen $\phi(x)$.

Several implementations of \vegas\ exist, e.g. Lepage's
\texttt{FORTRAN77} implementation \cite{Lepage:123074} or the GNU
scientific library C implementation \cite{gough2009gnu}.  Here, we
relate to the \vamp\ integration package as it is contained in \wz.
It provides an independent implementation which realizes the same
basic algorithm and combines it with multi-channel integration, as
explained below in Sec.~\ref{sec-3}.

Let us express the integration variables $x$ in terms of another set
of variables $r$, defined on the same unit
hypercube.  The mapping $r=G(x)$ is assumed bijective, factorizable, and
depends on a finite set of adjustable parameters.  If $r_i$ are
uniformly distributed random numbers, the distribution of $x_i=G^{-1}(r_i)$
becomes non-uniform, and we have to compensate for this by dividing by
the Jacobian $g(x)=G'(x)$,
\begin{equation}
  \label{eq:IOmega}
  I_\Omega[f] =
  \int_U f_\phi(x)\,\dd^dx
  =
  \int_{G(U)=U}
  \left.
    \frac{f_\phi(x)}{g(x)}
  \right|_{x=G^{-1}(r)}
  \dd^d r \; .
\end{equation}
Alternatively, we may interpret this result as the average of
$f_\phi/g$, sampled with an $x$ distribution that follows the
probability density $g(x)$, cf.~\cite{Ohl_1999}:
\begin{equation}
  \label{eq:1}
  I_\Omega[f]
  =
  \left\langle\frac{f}{g}\right\rangle_g
\end{equation}
For a finite sample with $N$ events, the estimators for integral and
variance are now given by
\begin{align}
  \label{eq:4}
  E_N(I)
  &=
    \frac{1}{N} \sum_{i=1}^N \frac{f_\phi(x_i)}{g(x_i)},
  \\
  V_N(I)
  &=
    \frac{N}{N - 1}
    \left(
    \frac{1}{N}\sum_{i=1}^N\left(\frac{f_\phi(x_i)}{g(x_i)}
    \right)^2
    - E_N(I)^2 \right),
\end{align}
where the $x_i$ are computed from the uniformly distributed $r_i$ via
$x_i=G^{-1}(r_i)$.

The \vegas\ algorithm makes the following particular choice for the
mapping $G$ (or its derivative $g=G'$):  For each dimension $k=1,\dots
d$, the interval $(0,1)$ is divided into $n_k$ bins $B_{kj_k}$ with
bin width $\Delta x_{kj_k}$, $j_k=1,\dots n_k$, such that the
one-dimensional probability distribution $g_k(x_k)$ is constant over
that bin $B_{kj_k}$ and equal to its inverse bin width, $1/\Delta
x_{kj_k}$. The overall probability distribution $g(x)$ is also
constant within each bin and given by
\begin{equation}
  \label{eq:g(x)}
  g(x) = \prod_k g_k(x_k) = \prod_k  \frac{1}{n_k\Delta x_{kj_k}},
\end{equation}
if $x_k\in B_{kj_k}$.  It is positive definite and satisfies
\begin{equation}
  \int_U g(x)\,\dd^d x = 1
\end{equation}
by construction, and thus defines an acceptable mapping $G$.

The iterative adaptation algorithm starts from equidistant bins.  It
consists of a sequence of integration passes, where after each pass,
the bin widths are adapted based on the variance distribution within
that pass~\cite{Peter_Lepage_1978,Lepage:123074}.
The adaptive mechanism of \vegas\ adjusts the size of the
bins for each bin $j_k$ of each axis $k$ based on the size of the
following measure:
\begin{equation}
  \label{eq:30}
  m_{j_k} =
  \left[ \left( \frac{\omega_{j_k}}{\sum \omega_{j_k}} - 1
    \right) \frac{1}{\log{\frac{\omega_{j_k}}{\sum \omega_{j_k}}}}
  \right]^\alpha \qquad
  \begin{dcases*}
    \quad > 1 & increase bin size \\
    \quad = 1 & keep bin size\\
    \quad < 1 & decrease bin size
  \end{dcases*} \qquad ,
\end{equation}
respecting the overall normalization.  The power $\alpha$ is a free
parameter.  The individual (squared) bin weights in Eq.~\eqref{eq:30}
are defined as
\begin{equation}
  \label{eq:31}
  \omega_{j_k}^2
  = \left( \langle f_{i} \rangle \Delta x_{i} \right)^2 =
  \sum_{\mathclap{\substack{i \\ (x_i)_k \in B_{kj_k}}}}
  \frac{f^2(x_i)}{g^2(x_i)}\,\rho(x_i).
\end{equation}
i.e., all integration dimensions $k'\neq k$ are averaged over when
adjusting the bins along dimension~$k$.  These bin weights are easily
accumulated while sampling events for the current integration pass.

This is an optimization problem with $n=\sum_{k=1}^d (n_k-1)$ free
parameters, together with a  specific strategy for optimization.  If
successful, the numerical variance of the ratio $f_\phi(x)/g(x)$ is
reduced after each adaptation of $g$.  In fact, the shape of $g(x)$ will
eventually resemble a histogrammed version of $f_\phi(x)$, with a
saw-like profile along each integration dimension.  Bins will narrow
along slopes of high variation in $f_\phi$, such that the ratio
$f_\phi/g$ becomes bounded.  The existence of such a bound is
essential for unweighting events, since the unweighting efficiency
$\epsilon$ scales with the absolute maximum of $f_\phi(x)/g(x)$ within
the integration domain. Clearly, the value of this maximum can only
be determined with some uncertainty since it relies on the finite
sample $\{x_i\}$. The saw-like shape puts further limits on the
achievable efficiency $\epsilon$.  Roughly speaking, each direction
with significant variation in $f_\phi$ reduces $\epsilon$ by a factor
of two. 

The set of updated parameters $\Delta x_{kj_k}$ defines the
integration grid for the next iteration.  In the particle-physics
applications covered by \wz\ we have $d\lesssim 30$, the number of
bins is typically chosen as $n_k\lesssim30$, all $n_k$ equal, so a
single grid consists of between a few and $10^3$ parameters $n$
subject to adaptation.  In practice, the optimization strategy turns
out to be rather successful.  Adapting the grid a few times does
actually improve the accuracy $a$ and the efficiency $\epsilon$
significantly.  Only the grids from later passes are used for
calculating observables and for event generation.  Clearly, the
achievable results are limited by the degree of factorizability of the
integrand.

\subsection{The \vegas\ algorithm: (pseudo-)stratified sampling}
\label{sec:strat-mode}

The importance-sampling method guarantees, for a fixed grid, that the
estimator $E_N$ for an integrand approaches the exact integral for
$N\to\infty$.  Likewise, a simulated unweighted event sample
statistically approaches an actual observed event sample, if the
integrand represents the actual matrix element.

However, the statistical distribution of the numbers $x_i$ is a rather
poor choice for an accurate estimate of the integral.  In fact, in one
dimension a simple equidistant bin-midpoint choice for $x_i$ typically
provides much better convergence than $1/\sqrt{N}$ for the random
distribution.  A reason for nevertheless choosing the Monte-Carlo
algorithm is the fact that for $n$ bins in $d$ dimensions, the total
number of cells is $n^d$, which easily exceeds realistic values for
$N$: for instance, $n=20$ and $d=10$ would imply $n^d=10^{13}$, but
evaluating the integrand at much more than $10^7$ points may already
become infeasible.

The \emph{stratified sampling} approach aims at combining the
advantages of both methods.  Binning along all coordinate axes
produces $n^d$ cells.  Within each cell, the integrand is evaluated at
precisely $s$ distinct points, $s\geq 2$.  We may choose $n$ such that
the total number of calls, $N=s \cdot n^d$, stays within limits
feasible for a realistic sampling. For instance, for $s=2$, $d=10$,
and limiting the number of calls to $N\approx 10^7$, we obtain
$n=4\dots 5$.  Within each cell, the points are randomly chosen,
according to a uniform distribution.  Again, the \vegas\ algorithm
iteratively adapts the binning in several passes, and thus improves
the final accuracy.

For the problems addressed by \wz, pure stratified sampling is not
necessarily an optimal approach.  The structure of typical integrands
cannot be approximated well by the probability distribution
$g(x)$ if the number of bins per dimension $n$ is small.  To allow
for larger $n$ despite the finite total number of calls, the
\emph{pseudo-stratified} approach applies stratification not in $x$
space, which is binned into $n_x^d$ cells with $n\lesssim 20$, but in
$r$ space which was not binned originally.  The $n_r$ bins in $r$ space are
not adapted, so this distribution stays uniform.  In essence, the
algorithm scans over all $n_r^d$ cells in $r$ space and selects two
points randomly within each $r$ cell, and then maps those points to
points in $x$ space, where they end up in any of the $n_x^d$ cells.
The overall probability distribution in $x$ is still $g(x)$ as given
by Eq.~\eqref{eq:g(x)}, but the distribution has reduced randomness in 
it and thus yields a more accurate integral estimate.

Regardless of the integration algorithm, simulation of unweighted
events can only proceed via strict importance sampling.  Quantum
mechanics dictates that events have to be distributed statistically
independent of each other over the complete phase space.
Therefore, \wz\ separates its workflow into integration passes which
adapt integration grids and evaluate the integral, and a subsequent
simulation run which produces an event sample.  The integration passes
may use either method, while event generation uses importance sampling
and, optionally, unweighting the generated events.  In practice, using
grids which have been optimized by stratified sampling is no
disadvantage for subsequent importance sampling since both
sampling methods lead to similarly shaped grids.

\subsection{Multi-channel integration}
\label{sec-3}

The adaptive Monte-Carlo integration algorithms described above do not yield
satisfactory results if the effective integrand $f_\phi$ fails to
factorize for the phase-space channel $\phi$.  In non-trivial
particle-physics processes, many different Feynman graphs, possibly with
narrow resonances, including mutual interference, contribute to the
integrand.

Ref.~\cite{Kleiss_1994} introduced a multi-channel ansatz for
integration that ameliorates this problem.  The basic idea is to
introduce a set of $K$ different phase-space channels $\phi_c:
U\to\Omega$, corresponding coordinates $x_c$ with $p=\phi_c(x_c)$ and
densities $\rho_\phi(x_c)$ with
$\dd\mu(p)=\phi'_c(x_c)\,\rho_\phi(x_c)\,\dd^d x_c$, and a
corresponding set of \emph{channel weights} $\alpha_c\in\Rset$ which
satisfy
\begin{equation}
  0\leq \alpha_c\leq 1,
  \qquad
  \sum_{c=1}^K\alpha_c = 1 \quad .
\end{equation}
We introduce the function
\begin{equation}
  \label{eq:h(p)}
  h(p) = \sum_c\alpha_c \frac{1}{\phi_c'(\phi_c^{-1}(p))} \quad , 
\end{equation}
which depends on the Jacobians $\phi'_c$ of all channels. Using this,
we construct a partition of unity,
\begin{equation}
  \label{eq:partition}
  1 = \sum_c\alpha_c\frac{1}{\phi'_c(\phi_c^{-1}(p))\,h(p)} \quad ,
\end{equation}
which smoothly separates phase space into regions where the
singularities dominate that are mapped out by any individual channel
$\phi_c$, respectively.

The master formula for multi-channel integration makes use of this
partition of unity and applies, for each term, its associated channel
mapping $\phi_c$.
\begin{align}
  I_\Omega[f]
  &=
    \int_\Omega f(p)\,\dd\mu(p)
    =
    \int_\Omega\sum_c\alpha_c\frac{f(p)}{h(p)}\,\frac{\dd\mu(p)}{\phi_c'}
  \\
  &=
    \sum_c\alpha_c
    \int_U
    \frac{f(\phi_c(x_c))}{h(\phi_c(x_c))}\,
    \rho_{\phi_c}(x_c)\,\dd^dx_c \quad .
\end{align}
The mappings $\phi_c$ are chosen such that any singularity of $f$ is
canceled by at least one of the Jacobians $\phi'_c$.  In the vicinity
of this singularity, $\phi'_c$ approaches zero in Eq.~\eqref{eq:h(p)},
and the effective integrand
\begin{equation}
  f^h_c(x_c)
  =
  \frac{f(\phi_c(x_c))}{h(\phi_c(x_c))}\,
  \rho_{\phi_c}(x_c),
\end{equation}
becomes
\begin{align}
  f^h_c(x_c)
  &\sim
    \frac{1}{\alpha_c} f_{\phi_c}(x_c)  \quad .
\end{align}
We thus benefit from the virtues of phase-space mapping in the
original single-channel version, but cancel all singularities at once.
Each effective integrand $f^h_c$ which depends on all weights
$\alpha_c$ and Jacobians $\phi'_c$ simultaneously, is to be integrated
in its associated phase space channel.  The results are added, each
integral weighted by $\alpha_c$.

The importance sampling method can then be applied as before,
\begin{equation}
  E_N[f]
  =
  \sum_c\alpha_c\frac{1}{N_c}\sum_{i_c=1}^{N_c} f^h_c(x_{ci_c}),
\end{equation}
where the total number of events $N$ is to be distributed among the
integration channels, $N=\sum_c N_c$.  A possible division is to
choose $\alpha_c$ such that $N_c=\alpha_c N$ is an integer for each
channel, and thus
\begin{equation}
  E_N[f]
  =
  \frac{1}{N}\sum_c\sum_{i_c=1}^{N_c} f^h_c(x_{ci_c})
\end{equation}
becomes a simple sum where the integration channels are switched
according to their respective weights.  Within each channel, the
points $x_{ci_c}$ can be taken as uniformly distributed random
numbers.  Alternatively, we may apply stratified sampling as before,
within each channel.

The weights $\alpha_c$ are free parameters, and thus an obvious
candidate for optimization.  We may start from a uniform distribution
of weights among channels, $\alpha_c = 1/K$, and adapt the weights
iteratively.  In analogy to the \vegas\ rebinning algorithm, we
accumulate the total variance for each channel $c$ to serve as a
number $\omega_c$ which enters an update formula analogous
to Eq.~\eqref{eq:31}, with an independent power~$\beta$ (cf. Eq.~(24)
in Ref.~\cite{Ohl_1999}, and Ref.~\cite{Kleiss_1994}):
\begin{equation}
  \label{eq:15}
  \alpha_c \to
  \frac{\alpha_c \omega_c^{\beta}}
  {\sum_c \alpha_c\omega_c^{\beta}} \quad .
\end{equation}
This results in updated weights $\alpha_c$.  The weights, and the
total number of events $N$ for the next iteration, are further
adjusted slightly such that the event numbers $N_c$ become again integer.
Furthermore, it may be useful to insert safeguards for channels which
by this algorithm would acquire very low numbers $N_c$, causing
irregular statistical fluctuations.  In that case, we may
choose to either switch off such a channel, $\alpha_c=0$, or keep
$N_c$ at some lower threshold value, say $N_c=10$.  These refinements
are part of the \wz\ setup.

Regarding particle-physics applications, a straightforward translation
of (archetypical representatives of) Feynman graphs into integration
channels can result in large values for the number of channels $K$, of
order $10^5$ or more.  In fact, if the number of channels increases
proportional to the number of Feynman graphs, it scales factorially
with the number of elementary particles in the process.  This is to be
confronted with the complexity of the transition-matrix calculation,
where recursive evaluation results in a power law.  Applied naively,
multi-channel phase-space sampling can consume the dominant fraction
of computing time.  Furthermore, if the multi-channel approach is
combined with adaptive binning (see below), the number of channels is
multiplied by the number of grid parameters, so the total number of
parameters grows even more quickly.  For these reasons, \wz\ contains
a heuristic algorithm that selects a smaller set of presumably
dominant channels for the multi-channel integration. Since all
parameterizations are asymptotically equivalent to each other
regarding importance sampling, any such choice does not affect 
the limit $E_N[f]\to I_\Omega[f]$.  It does affect the variance and
can thus speed up -- or slow down -- the convergence of the integral
estimates for $N\to\infty$ and for iterative weight adaptation.

\subsection{Doubly adaptive multi-channel integration: \vamp}

The \vamp\ algorithm combines multi-channel integration with channel
mappings $\phi_c$ with the \vegas\ algorithm.  For each channel
$c=1,\dots K$, we introduce a bijective step mapping $G_c$ of the unit
hypercube onto itself $U\to G_c(U)=U$.  The Jacobian is
$g_c(x)=G'_c(x)$, where $g_c$ factorizes along the coordinate axes
(labeled by $k=1,\dots d$) and is constant within bins $B_{c,kj_k}$
(labeled by $j_k=1,\dots n_{c,k}$),
\begin{equation}
  \label{eq:gc(x)}
  g_c(x)
  =
  \prod_k g_{c,k}(x_{c,k}) = \prod_k  \frac{1}{n_{c,k}\Delta
    x_{c,kj_k}} \; ,
\end{equation}
if $x_{c,k}\in B_{c,kj_k}$.  The normalization condition
\begin{equation}
  \int_U g_c(x_c)\,\dd^d x_c = 1
\end{equation}
is satisfied for all channels, and enables us to construct $G_c$.

We chain the mappings $G_c$ with the channel mappings $\phi_c$ in the
partion of unity, Eq.~\eqref{eq:partition}, and write
\begin{equation}
  \label{eq:vamp-master}
  I_\Omega[f]
  =
  \left.
    \sum_c\alpha_c\int_U
    f^g_c(x_c)
  \right|_{x_c = G_c^{-1}(r_c)} 
  \dd^d r_c \quad ,
\end{equation}
where the modified effective integrand for channel $c$ is given by
\begin{equation}
  f^g_c(x_c)
  =
  \frac{f(\phi_c(x_c))}{g(\phi_c(x_c))} \rho_{\phi_c}(x_c) \quad .
\end{equation}
Here, $g(p)$ replaces $h(p)$ in~\eqref{eq:h(p)},
\begin{equation}
  \label{eq:g(p)}
  g(p)
  =
  \left.
    \sum_c\alpha_c \frac{g_c(x_c)}{\phi_c'(x_c)}
  \right|_{x_c = \phi_c(p)}  \quad .
\end{equation}
The variance of the integrand is reduced not just by the fixed
Jacobian functions $\phi'_c$, but also by the tunable distributions
$g_c$.  In a region where one of the $g_c$ distributions becomes
numerically dominant, $\alpha_c f^g_c(x_c)$ approaches the
single-channel expression $f_{\phi_c}(x_c)/g_c(x_c)$,
cf.~Eq.~\eqref{eq:IOmega}.

The multi-channel sampling algorithm can be expressed in
form of the integral estimate $E_N[f]$,
\begin{equation}
  E_N[f]
  =
  \frac{1}{N}\sum_c\sum_{i_c=1}^{N_c} f^g_c(x_{ci_c})
  \qquad
  \text{with}
  \qquad
  x_{ci_c} = G_c^{-1}(r_{ci_c}),
\end{equation}
and the $r_{ci_c}$ are determined either from a uniform probability
distribution in the unit hypercube, or alternatively, from a uniform
probability distribution within each cell of a super-imposed
stratification grid.  The free parameters of this formula are
$\alpha_c$, $c=1,\dots K$, and for each channel,
the respective channel grid with parameters $\Delta x_{c,k,j_k}$.

In order to improve the adaptation itself, similarity mappings between
different channels can be used in order to achieve a better adaptation
of the individual grids, because the distribution for grid adaptation
is better filled. This in turn leads to an improved convergence of the
integration and a better weighting efficiency of the event
generation. For this, we define equivalences for channels that share a
common structure. These equivalences allow adaptation information,
i.e.~the individual bin weights $w_{j_k}$ of each axis, to be averaged
over several channels, which in turn improves the statistics of the
adaptation. Such an equivalence maps the individual bin weights of a
channel $c^{\prime}$ onto the current channel $c$ together with the
permutation of the integration dimensions $d$, $\pi: c \mapsto
c^{\prime} : k \mapsto \pi(k)$, and the type of mapping.
The different mappings used in that algorithm are:
\begin{equation}
  \begin{array}{ll}
    \text{{\bf identity}}  &
    \mbox{}\qquad
    w^{c}_{j_k} \rightarrow w^{c}_{j_k} +
    w^{c^{\prime}}_{j_{\pi(k)}} \\
    \text{{\bf invert}}  &
    \mbox{} \qquad
    w^{c}_{j_k} \rightarrow w^{c}_{j_k} +
    w^{c^{\prime}}_{d - j_{\pi(k)}}
    \\
    \text{{\bf symmetric}} &
    \mbox{} \qquad
    w^{c}_{j_k} \rightarrow w^{c}_{j_k} + \frac12
    (w^{c^{\prime}}_{j_{\pi(k)}} + w^{c^{\prime}}_{d - j_{\pi(k)}})
    \\[5pt]
    \text{{\bf invariant}} &
    \mbox{} \qquad
    w^{c}_{j_k} \rightarrow 1  
  \end{array}
\end{equation}
Interesting applications are those that require very high statistics,
as shown in the table~\ref{tab:comparison}. Channel equivalences have
been shown to play a crucial role in sampling correctly particularly
the less densely populated regions of phase space, e.g. in vector
boson scattering~\cite{Beyer:2006hx,Alboteanu:2008my,Kilian:2014zja,
Kilian:2015opv,Fleper:2016frz,Ballestrero:2018anz,Brass:2018hfw}
or in beyond the Standard Model (BSM) simulations with a huge number
of phase space channels~\cite{Hagiwara:2005wg,Reuter:2010nx,
Pietsch:2012nu,Reuter:2012ng}. Channel equivalences are used both for
the traditional \vamp\ Monte Carlo integrator as well as for the new
parallelized version. Constructing these channel equivalences
is part of the phase space algorithm, but a detailed explanation is
out of scope of this work.

\begin{table}[h]
  \centering
  \sisetup{
  scientific-notation=fixed,
  separate-uncertainty=true,
}
\begin{tabular}{
  l
  S %[table-format=1.8e+1]
  S %[table-format=1.8e+1]
  }
  \toprule
  Process & {$\sigma_{\text{new}} \SI{}{\per\femto\barn}$} & {$\sigma_{\text{original}} \SI{}{\per\femto\barn}$} \\
  \midrule

  $\Pgluon \Pgluon \to \PW \Pquark \APquark$ & 24097(22) & 24091(12) \\

  $\Pgluon \Pgluon \to \PW \Pquark \APquark \Pgluon$ & 9111(15) & 9142(6) \\

  $\Pgluon \Pgluon \to \PW \Pquark \APquark \Pgluon \Pgluon$ & 2.31(7)e+03 & 2363(7) \\

  $\Pgluon \Pgluon \to \PW \Pquark \APquark \Pgluon\Pgluon \Pgluon$ & 410(130) & 523(6) \\

  $\Pjet\Pjet \to \PW \Pjet$ & 9.366(6)e+05 & 9.36(28)e+05 \\

  $\Pjet\Pjet \to \PW \Pjet \Pjet$ & 2.8749(26)e+05 & 2.87(14)e+05 \\

  $\Pjet\Pjet \to \PW \Pjet \Pjet \Pjet$ & 7.957(15)e+04 & 8.0(6)e+04 \\

  $\Pjet\Pjet \to \PW \Pjet \Pjet \Pjet \Pjet$ & 1.79(11)e+04 & 2.09(6)e+04 \\

  \bottomrule
\end{tabular}

  \caption{We show the numerical results for the cross sections of
    $\PW + \text{jets}$ processes computed with the original \vamp and
    the reimplemented version. Both process and cut definitions have
    been taken from \cite{Kilian:2007gr}, as well as the original
    results of \vamp. The original results were calculated with
    equivalences and the results of the new implementation without
    equivalences. By comparison, the results are statistically
    consistent, but the results without equivalences are less
    accurate. 
  }
  \label{tab:comparison}
\end{table}

%%%%% End Equivalences

The actual integration algorithm is organized as follows.  Initially,
all channel weights and bin widths are set equal to each other.  There
is a sequence of iterations where each step consists of first generating
a sample of $N$ events, then adapting the free parameters.  This
adaptation may update either the channel weights via Eq.~\eqref{eq:15}
or the grids via Eq.~\eqref{eq:30}, or both, depending on user
settings. The event sample is divided among the selected channels
based on event numbers $N_c$.  For each channel, the integration
hypercube in $r$ is scanned by cells in terms of stratified sampling,
or sampled uniformly (importance sampling).  For each point $r_c$, we
compute the mapped point $x_c$, the distribution value $g_c(x_c)$, and
the phase-space density $\rho_c(x_c)$ at this point.  Given the fixed
mapping $\phi_c$, we compute the phase-space point $p$ and the
Jacobian factor $\phi'_c$.  This allows us to evaluate the integrand
$f(p)$.  Using $p$, we scan over all other channels $c'\neq c$ and
invert the mappings to obtain $\phi'_{c'}$, $x_{c'}$,
$g_{c'}(x_{c'})$, and $\rho_{c'}(x_{c'})$.  Combining everything, we
arrive at the effective weight $w=f_c^g(x_c)$ for this event.
Accumulating events and evaluating mean, variance, and other
quantities then proceeds as usual.  Finally, we may combine one or
more final iterations to obtain the best estimate for the integral,
together with the corresponding error estimate.

If an (optionally unweighted) event sample is requested, \wz\ will take
the last grid from the iterations and sample further events, using the
same multi-channel formulas, with fixed parameters, but reverting to
importance sampling over the complete phase space.  The channel
selection is then randomized over the channel weights $\alpha_c$,
allowing for an arbitrary number of simulated physical events.

%%%%%

\section{Parallelization of the \wz\ workflow}
\label{sec:parallel}

In this section we discuss the parallelization of the \wz\ integration
and event generation algorithms. We start with a short definition of
observables and timings that allow to quantify the gain of a
parallelization algorithm in Sec.~\ref{sec:parallel_basics}. Then, in
Sec.~\ref{sec:tasks_whizard}, we discuss the computing tasks for a
typical integration and event generation run with the \wz\ program,
while in Sec.~\ref{sec:parallel_tools} we list possible computing
frameworks for our parallelization tasks and what we chose to
implement in \wz. Random numbers have to be set up with great care for
parallel computations, as we point out in
Sec.~\ref{sec:random_parallel}. The \wz\ algorithm for parallelized
integration and event generation is presented in all details in
Sec.~\ref{sec:whizard_parallel}. Finally, in Sec.~\ref{sec:wood2}, we
introduce an alternativ method to generate the phase-space
parameterization that is more efficient for higher final-state
particle multiplicities and is better suited for parallelization. 

\subsection{Basics}
\label{sec:parallel_basics}

The time that is required for a certain computing task can be reduced
by employing not a single processing unit (\emph{worker}), but several workers which
are capable of performing calculations independent of each other.  In
a slightly simplified view, we may assume that a bare program consists
of parts that are performed by a single worker (time $T_s$), and of other
parts that are performed by $n$ workers simultaneously (time $T_m$).  The
serial time $T_s$ also covers code that is executed identically on all
workers.  The total computing time can then be written as
\begin{equation}
  T(n) = T_s + \frac{1}{n}T_m + T_c(n).
\end{equation}
The extra time $T_c$ denotes the time required for communication
between the workers, and for workers being blocked by waiting conditions.  Its
dependence on $n$ varies with the used algorithm, but we expect a
function that vanishes for $n=1$ and monotonically increases with
$n$, e.g., $T_c\sim\log(n)$ or $T_c\sim (n-1)^\alpha$, $\alpha>0$.  The
\emph{speedup} factor of parallelization then takes the form
\begin{equation}
  \label{eq:speedup}
  f(n) = \frac{T(1)}{T(n)}
  = \frac{T_s + T_m}{T_s + \frac{1}{n}T_m + T_c(n)} \quad ,
\end{equation}
and we want this quantity to become as large as possible.

Ideally, $T_s$ and $T_c$ vanish, and
\begin{equation}
  f(n) = n \quad ,
\end{equation}
but in practice the serial and communication parts limit this
behavior.  As long as communication can be neglected, $f(n)$
approaches a plateau which is determined by $T_s$,
\begin{equation}
  f(n) \leq 1 + \frac{T_m}{T_s} \quad .
\end{equation}
Eventually, the communication time $T_c(n)$ starts to dominate and
suppresses the achievable speedup again,
\begin{equation}
  f(n) \to \frac{T_s+T_m}{T_c(n)} \to 0 \;.
\end{equation}
Clearly, this behavior limits the number $n$ of workers that we can
efficiently employ for a given task.

The challenges of parallelization are thus twofold: (i) increase the
fraction $T_m/T_s$ by parallelizing all computing-intensive parts of
the program.  For instance, if $T_s$ amounts to $0.1\,\%$ of $T_m$,
the plateau is reached for $n=1000$ workers.  (ii) make sure that at
this saturation point, $T_c(n)$ is still negligible.  This can be
achieved by (a) choose a communication algorithm where $T_c$ increases
with a low power of $n$, or (b) reduce the prefactor in $T_c(n)$,
which summarizes the absolute amount of communication and blocking per
node.

We will later on in the benchmarking of our parallelization algorithm
compare the speedup to Amdahl's law~\cite{Amdahl_1967}. For this, we
neglect the communication time in Eq.~\eqref{eq:speedup}, and write
the time executed by the parallelizable part as a fraction $p \cdot T$
of the total time $T = T_s + T_M$ of the serial process, while the
non-parallelizable part is then $(1-p)T$. The speedup in
Eq.~\eqref{eq:speedup} translates then into
\begin{equation}
  f(n) = \frac{1}{(1-p) + \frac{p}{n}} \;\; \overset{n \to
    \infty}{\longrightarrow} \;\; \frac{1}{1-p} \qquad.
\end{equation}
In Sec.~\ref{sec:speedup} we use Amdahl's law as a comparison for
parallelizable parts of $p= 90 \%$ and 100 \%, respectively. Note that
Amdahl's law is considered to be very critical on the possible
speedup, while a more optimistic or realistic estimate is given by
Gustafson's law~\cite{Gustafson_1988}. To discuss the differences,
however, is beyond the scope of this paper.

\subsection{Computing tasks in \wz}
\label{sec:tasks_whizard}

The computing tasks performed by \wz\ vary, and crucially depend on
the type and complexity of the involved physics processes.  They also
depend on the nature of the problem, such as whether it involves
parton distributions or beam spectra, the generation of event files,
or scans over certain parameters, or whether it is a LO or NLO
process.

To begin with, we therefore identify the major parts of the program
and break them down into sections which, in principle, can contribute
to either $T_s$ (serial), $T_m$ (parallel), or $T_c$ (communication).

\paragraph{\sindarin.}
All user input is expressed in terms of \sindarin\ expressions, usually
collected in an input file.  Interpreting the script involves
pre-processing which partly could be done in parallel.  However,
the \sindarin\ language structure allows for mixing declarations with
calculation, so parallel pre-processing can introduce nontrivial
communication.  Since scripts are typically short anyway,
we have not yet considered parallel evaluation in this area.  This
also applies for auxiliary calculations that are performed within
\sindarin\ expressions.

\paragraph{Models.}
Processing model definitions is done by programs external to the
\wz\ core in advance. We do not consider this as part of the
\wz\ workflow.  Regarding reading and parsing the resulting model
files by \wz, the same considerations apply as for the
\sindarin\ input.  Nevertheless, for complicated models such as the
MSSM, the internal handling of model data involves lookup tables.  In
principle, there is room for parallel evaluation.  This fact has not
been exploited, so far, since it did not constitute a bottleneck.

\paragraph{Process construction.}
Process construction with \wz, i.e., setting up data structures that
enable matrix-element evaluation, is delegated to programs external to
the \wz\ core. For tree-level matrix elements, the in-house
\om\ generator constructs Fortran code which is compiled and linked to
the main program.  For loop matrix elements, \wz\ relies on programs
such as \go, \re, or \ol. The parallelization capabilities rely on
those extra programs, and are currently absent.  Therefore,
process-construction time contributes to $T_s$ only.

\paragraph{Phase-space construction.}
Up to \wz\ version 2.6.1, phase-space construction is performed
internally with \wz\ (i.e. by the \wz\ core), by a module which
recursively constructs data structures derived from simplified Feynman
graphs.  The algorithm is recursive and does not lead to obvious
parallelization methods; the resulting $T_s$ contribution is one of
the limiting factors.

A new algorithm, which is described below in Sec.~\ref{sec:wood2},
re-uses the data structures from process construction via \om.  The
current implementation is again serial ($T_s$), but significantly more
efficient.  Furthermore, since it does not involve recursion it can be
parallelized if the need arises.

\paragraph{Integration.}
Integrating over phase space involves the \vamp\ algorithm as
described above.  In many applications, namely those with complicated
multi-particle or NLO matrix elements, integration time dominates the
total computing time.  We can identify tasks that qualify for serial,
parallel, and communication parts:
\begin{itemize}
\item
  Initialization.  This part involves serial execution.  If subsequent
  calculations are done in parallel, it also involves communication,
  once per process.
\item
  Random-number generation.  The \vamp\ integrator relies on
  random-number sequences.  If we want parallel evaluation on separate
  workers, the random-number generator should produce independent,
  reproducible sequences without the necessity for communication or blocking.
\item
  \vamp\ sampling.  Separate sampling points involve independent
  calculation, thus this is a preferred target for turning serial
  into parallel evaluation.  The (pseudo-) stratified algorithm involves some
  management, therefore communication may not be entirely avoidable.
\item
  Phase-space kinematics.  Multi-channel phase-space evaluation
  involves two steps: (i) computing the mapping from the unit
  hypercube to a momentum configuration, for a single selected
  phase-space channel, and (ii) computing the inverse mapping, for all
  other phase-space channels.  The latter part is a candidate for
  parallel evaluation.  The communication part involves distributing
  the momentum configuration for a single event.  The same
  algorithm structure applies to the analogous discrete mappings
  introduced by the \vamp\ algorithm.
\item
  Structure functions.  The external PDF library for hadron collisions
  (\textsc{Lhapdf}) does not support intrinsic parallel evaluation.
  This also holds true for the in-house \textsc{Circe1}/\textsc{Circe2} 
  beamstrahlung library.
\item
  Matrix-element evaluation.  This involves sums over quantum numbers:
  helicity, color, and particle flavor.  These sums may be distributed
  among workers.  The tradeoff of parallel evaluation has to be weighted
  against the resulting communication.  In particular, common
  subexpression elimination or caching partial results do optimize
  serial evaluation, but actually can inhibit parallel evaluation or
  introduce significant extra communication.
\item
  Grid adaptation.  For a grid adaptation step, results from all sampling points
  within a given iteration have to be collected, and the adapted grids
  have to be sent to the workers.  Depending on how grids are distributed,
  this involves significant communication.  The calculations for
  adapting grids consume serial time, which in principle could also be
  distributed.
\item
  Integration results.  Collecting those is essentially a byproduct of
  adaptation, and thus does not involve extra overhead.
\end{itemize}

\paragraph{Simulation.}
A simulation pass is similar to an integration pass.  There is no grid
adaptation involved.  The other differences are
\begin{itemize}
\item
  Sampling is done in form of strict importance sampling.  This is
  actually simpler than sampling for integration.
\item
  Events are further transformed or analyzed.  This involves simple kinematic
  manipulations, or complex calculations such as parton shower and
  hadronization.  The modules that are used for such tasks, such as
  \pythia\ or \wz's internal module, do not support intrinsic
  parallelization.  Generating histograms and plots involves
  communication and some serial evaluation. 
\item
  Events are written to file.  This involves communication and serial
  evaluation, either event by event, or by combining event files
  generated by distinct workers.
\end{itemize}

\paragraph{Rescanning events.}
In essence, this is equivalent to simulation.  The difference is that
the input is taken from an existing event file, which is scanned
serially.  If the event handling is to be distributed, there is
additional communication effort.

\paragraph{Parameter scans.}
Evaluating the same process(es) for different sets of input data, can
be done by scripting a loop outside of \wz.  In that case,
communication time merely consists of distributing the input
once, and collecting the output, e.g., fill plots or histograms.  However, there are
also contributions to $T_s$, such as compile time for process code.
Alternatively, scans can be performed using \sindarin\ loop
constructs.  Such loops may be run in parallel.  This avoids some of
the $T_s$ overhead, but requires communicating \wz\ internal data
structures. Phase-space construction may contribute to either $T_s$ or
$T_m$, depending on which input differs between workers.  Process
construction and evaluation essentially turns into $T_m$.  This
potential has not been raised yet, but may be in a future extension.
The benefit would apply mainly for simple processes where the current
parallel evaluation methods are not efficient due to a small $T_m$
fraction.

\subsection{Paradigms and tools for parallel evaluation}
\label{sec:parallel_tools}

There are a number of well-established protocols for parallel
evaluation.  They differ in their overall strategy, level of language
support, and hardware dependence.  In the following, we list some
widely used methods.
\begin{enumerate}
\item
  MPI (message-passing interface, cf.~e.g.~\cite{Gropp_MPI}).  This
  protocol introduces a set of independent abstract workers which
  virtually do not share any resources.  By default, the program runs
  on all workers simultaneously, with different data sets.  (Newer
  versions of the protocol enable dynamic management of workers.)
  Data must be communicated explicitly via virtual buffers.  With
  MPI~3, this communication can be set up asynchronously and
  non-blocking.  The MPI protocol is well suited for computing
  clusters with many independent, but interconnected CPUs with local
  memory.  On such hardware, communication time cannot be neglected.

  For \fortran, MPI is available in form of a library, combined with a
  special run manager.
\item
  OpenMP (open multi-processing,
  cf.~e.g.~\cite{Chandra:2001:PPO:355074}).  This protocol assumes a
  common address space for data, which can be marked as local to
  workers if desired.  There is no explicit communication.  Instead,
  data-exchange constraints must be explicitly implemented in form of
  synchronization walls during execution.  OpenMP thus maps the
  configuration of a shared-memory multi-core machine.  We observe
  that with such hardware setup, communication time need not exceed
  ordinary memory lookup.  On the other hand, parallel
  execution in a shared-memory (and shared-cache) environment can run
  into blocking issues.

  \fortran\ compilers support
  OpenMP natively, realized via standardized preprocessor directives.
\item
  Coarrays (cf.~e.g.~\cite{CAF}).  This is a recent native
  \fortran\ feature, introduced in the \ffortran\ standard.  The
  coarray feature combines semantics both from MPI and OpenMP, in the
  form that workers are independent both in execution and in data, but
  upon request data can be tagged as mutually addressable.  Such
  addressing implicitly involves communication. 
\item
  Multithreading.  This is typically an operating-system feature which
  can be accessed by application programs.  Distinct threads are
  independent, and communication has to be managed by the operating
  system and kernel.
\end{enumerate}
The current strategy for parallel evaluation with \wz\ involves MPI
and OpenMP, either separately or in combination.  We do not use
coarrays, which is a new feature that did not get sufficient compiler
support, yet~\footnote{There would of course be the possibility to
  have a special version of \wz\ only available for the newest
  version(s) of compilers to test those features. This is part of a
  future project.}.  On the other hand, operating system threads are rather
unwieldy to manage, and largely superseded by the OpenMP or MPI protocols
which provide abstract system-independent interfaces to this functionality.

\subsection{Random numbers and parallelization}
\label{sec:random_parallel}

\wz uses pseudo random numbers to generate events. Most random number
generators have in common that they compute a reproducible 
stream of uniformly distributed random numbers $\{x_i\} \in (0, 1)$
from a given starting point (seed) and they have a relative large
periodicity. In addition, the generated random numbers should not have
any common structures or global correlations. To ensure these
prerequisites different test suites exist based on statistical 
principles and other methods. One is the \textsc{TestU01} library
implemented in \texttt{ANSI C} which contains several tests for
empirical randomness~\cite{lecuyer07_testu0}. A very extensive
collection of tests is the \textsc{Die Hard} suite~\cite{diehard},
also known as \textsc{Die Hard 1}, which contains e.g. the squeeze
test, the overlapping sums test, the parking lot test, the craps test,
and the runs test~\cite{wald1940}. There is also a more modern version
of this test suite, \textsc{Die Harder} or \textsc{Die Hard
2}~\cite{dieharder} which contains e.g. the
\textsc{Knuthran}~\cite{Knuth:1997:ACP:270146} and the 
\textsc{Ranlux}~\cite{Luscher:1993dy,Shchur:1998wi}
tests. Furthermore, the computation of the pseudo random numbers
should add as less as possible computation time.

The default random number generator of WHIZARD is the TAO random
number generator proposed by \cite{Knuth:1997:ACP:270146} and provided
by \vamp. This generator passes the \textsc{Die hard} tests. It is
based on a lagged Fibonacci sequence,
\begin{equation}
  \label{eq:fibonacci}
  X_{n+1} = \left( X_{n-k} + X_{n-l} \right) \mod 2^{30},
\end{equation}
with lags $k = 100$ and $l = 37$ computing portable, 30-bit integer
numbers. The computation needs a reservoir of random numbers of at
least $k = 100$ which have to be prepared in advance. To ensure a
higher computation efficiency, a reservoir of more than $1000$ is
needed. Furthermore, the TAO random numbers suffers from its integer
arithmetic. In general on modern CPUs floating point arithmetic is
faster and can be put in pipelines allowing terser computation. 

In order to utilize the TAO generator for a parallelized application
we have to either communicate each random number before or during
sampling, both are expensive on time, or we have to prepare or, at
least, guarantee independent streams of random numbers from different
instances of TAO by initializing each sequence with different seeds. 
The latter is hardly feasible or even impossible to ensure for all
combinations of seeds and number of workers. This and the
(time-)restricted integer arithmetic render the TAO random number
generator impractical for our parallelization task. 

To secure independent (and still reproducible) random numbers during
parallel sampling we have implemented the RNGstream algorithm by
\cite{L_Ecuyer_2002}. The underlying generator is a combined
multiple-recursive generator, referred to as \texttt{MRG32k3a}, based
on two multiple-recursive generators, 
\begin{align}
  \label{eq:mrg32k3a} x_{1, n} = \left( 1403580 x_{1, n-2} - 810728 x_{1,
  n-3}\right) \mod 4294967087,\\ x_{2, n} = \left( 527612 x_{2, n-1} - 1370589
  x_{2,n-3} \right) \mod 4294944443,
\end{align}
at the n-th step of the recursion with the initial seed $\vec{x}_{i, 0} =
(x_{i,-2}, x_{i, -1}, x_{i, 0})^{T}, i \in \{1, 2\}$.  The two states are then
combined to produce a uniformly-distributed random number $u_n$ as
\begin{align}
  \label{eq:rngstream}
  z_n = \left( x_{1, n} - x_{2, n} \right) \mod 4292967087, \\
  u_n =
  \begin{cases}
    z_n / 4294967088 & \text{if } z_n > 0. \\
    4292967087 / 4292967088 & \text{if } z_n = 0.
  \end{cases}
\end{align}
The resulting random number generator has a period of length $\approx
2^{191}$. It passes all tests of \textsc{TestU01} and \textsc{Die
Hard}. 

The overall sequence of random numbers is divided into streams of
length $2^{127}$, each of these streams is then further subdivided
into substreams of length $2^{76}$. Each stream or subsequent
substream can be accessed by repeated application of the transition
function $x_{n} = T(x_{n - 1})$. We rewrite the transition function as
a matrix multiplication on a vector, making the linear dependence
clear, $x_{n} = T \times x_{n - 1}$. Using the power of modular
arithmetic, the repated application of the transition function can be
precomputed and be stored making access of the (sub-)streams 
as simple as sampling one. In the context of the parallel evaluation of
the random number generator we can get either independent streams of
random numbers for each worker, or, conserving the numerical
properties for the integration process, assign each channel a stream
and each stratification cell of the integration grid a substream in
the serial and parallel run. Then we can easily distribute the workers
among channels and cells without further concern about the random
numbers. 

The original implementation of the RNGstream was in \texttt{C++} using
floating point arithmetic. We have rewritten the implementation for 
\ffortran\ in \wz. 

\subsection{Parallel evaluation in \wz}
\label{sec:whizard_parallel}

To devise a strategy for parallel evaluation, we have analyzed the
workflow and the scaling laws for different parts of the code, as
described above.  Complex multi-particle processes are the prime
target of efficient evaluation.  In general, such processes involve a
large number of integration dimensions, a significant number of
quantum-number configurations to be summed over, a large number of
phase-space points per iteration of the integration procedure, and a
large number of phase-space channels.  By contrast, for a single
phase-space channel the number of phase-space points remains
moderate.  

After the integration passes are completed, event generation in the simulation
pass is another candidate for parallel execution.  Again, a large number of
phase-space points have to be sampled within the same computational model as
during integration.  Out of the generated sample of partonic events, in the
unweighted mode, only a small fraction is further processed.  The subsequent
steps of parton shower, hadronization, decays, and file output come with their
own issues of computing (in-)efficiency.

We address the potential for parallel evaluation by two independent protocols,
OpenMP and MPI.  Both frameworks may be switched on or off independent of each
other.

\subsubsection{Sampling with OpenMP}

On a low-level scale, we have implemented OpenMP as a protocol for parallel
evaluation.  The OpenMP paradigm is intended to distribute workers among the
physical core of a single computing node, where actual memory is shared
between cores.  While in principle, the number of workers can be set
freely by the user of the code, one does expect improvements as long
as the number of workers is less or equal to the number of physical
cores.  The number of OpenMP workers therefore is typically between 1
and 8 for standard hardware, and can be somewhat larger for
specialized hardware.

We apply OpenMP parallelization for the purpose of running simple
\fortran\ loops in parallel.  These are
\begin{enumerate}
\item
  The loop over helicities in the tree-level matrix-element code that is generated
  by \om.  For a typical $2\to 6$ fermion process, the number of
  helicity combinations is $2^8=64$ and thus fits the expected number
  of OpenMP workers.  We do not parallelize the sum of the flavor or
  color quantum numbers.  In the current model of \om\ code, those
  sums are subject to common-subexpression elimination which inhibits
  trivial parallelization.
\item
  The loop over channels in the inverse mapping between phase-space
  parameters and momenta.  Due to the
  large number of channels, the benefit is obvious, while the
  communication is minimal, and in any case is not a problem in a
  shared-memory setup.
\item
  Analogously, the loop over channels in the discrete inverse
  mapping of the phase-space parameters within the \vamp\ algorithm.
\end{enumerate}
In fact, these loops cover the most computing-intensive tasks.  As
long as the number of OpenMP workers is limited, there is no
substantial benefit from parallelizing larger portions of code at this 
stage.

\subsubsection{Sampling with MPI}

The MPI protocol is designed for computing clusters. We will give a
short introduction into the terminology and the development of its
different standards over time in the next subsection. The MPI model
assumes that memory is local to each node, so data have to be
explicitly transferred between nodes if sharing is required. The
number of nodes can become very large. In practice, for a given
problem, the degree of parallelization and the amount of communication
limits the number of nodes where the approach is still practical. For
\wz, we apply the MPI protocol at a more coarse-grained level than
OpenMP, namely the loop over sampling points which is controlled by
the \vamp\ algorithm. 

As discussed above, in general, for standard multi-particle problems 
the number of phase-space channels is rather large, typically
exceeding $10^3\dots  10^4$. In that case, we assign one separate
channel or one subset of channels to each worker. In some
calculations, the matrix element is computing-intensive but the number
of phase-space channels is small (e.g. NNLO virtual matrix elements),
so this model does not apply. In that case, we parallelize over single
grids. We assign to each worker a separate slice of the $n_{r}^{d}$
cells of the stratification space. In principle, for the simplest case
of $n_{r} = 2$, we can exploit up to $2^{d}$ computing nodes for a
single grid. On the other hand, parallelization over the $r$-space is
only meaningful \textit{when} $n_{r} \geq 2$. Especially, when we take
into account that $n_{r}$ changes between different iterations as the
number of calls $N_C$ depends on the multi-channel weights
$\alpha_i$. Hence, we implement a sort of auto-balancing in the form
that we choose between the two modes of parallelization before and
during sampling in order to handle the different scenarios
accordingly. Per default, we parallelize over phase-space 
multi-channel, but prefer single-grid parallelization for the case
that the number of cells in $r$-space is $n_{r} > 1$. Because the
single-grid parallelization is finer grained than the phase-space
channel parallelization, this allows in principle to exploit more
workers. Furthermore, we note that the Monte Carlo integration itself
does not exhibit any boundary conditions demanding communication
during sampling (except when we impose such a condition by hand). In
particular, there is no need to communicate random numbers. We discuss
the details of the implementation later on.

\subsubsection{The Message-Passing Interface (MPI) Standard}

We give a short introduction into the terminology necessary to
describe our implementation below, and also into the message-passing
interface (MPI) standard. The MPI standard specifies a large 
amount of procedures, types, memory and process management and
handlers, for different purposes. The wide range of functionality
obscures a clear view on the problem of parallelization and on the
other hand it unnecessarily complicates the problem itself. So, we
limit the use of functionality to an absolute minimum. E.g., we do not
make use of the MPI shared-memory model and, for the time being, the
use of an own process management for a server-client model. In the
following we introduce the most common terms. In the implementation
details below, we again refer to the MPI processes as \textit{workers}
in order to not confuse them with the \wz's physical processes. 

The standard specifies MPI programs to consist of autonomous
processes, each in principle running its own code, in an
MIMD\footnote{Multiple instructions, multiple data. Machines
supporting MIMD have a number of processes running asynchronously
and independently.} style, cf.~\cite{MPI_3.1}, p. 20.  In order to
abstract the underlying hardware and to allow separate communication
between different program parts or libraries, the standard generalizes
as \textit{communicators} processes apart from the underlying hardware
and introduces communication contexts. Context-based communication
secures that different messages are always received and sent within
their context, and do not interfere with messages in other contexts. 
Inside communicators, processes are grouped (process group) as ordered
sets with assigned ranks (labels) ${0, \dots, n -1}$. The predefined
\verb|MPI_COMM_WORLD| communicator contains all processes known at the
initialization of a MPI program in a linearly ordered fashion. In 
most cases, the linear order does not reflect the architecture of the
underlying hardware and network infrastructure, therefore, the
standard defines the possibility to map the processes onto the
hardware and network infrastructure to optimize the usage of resources
and increase the speedup.

A way to conceivably optimize the parallelization via MPI is to make
the MPI framework aware about the communication flows in your
application. In the group of processes in a communicator, not all
processes will communicate with every other process. The network of
inter-process communication is known as MPI \emph{topologies}. The
default is \verb|MPI_UNDEFINED| where not specific topology has been
specified, while \verb|MPI_CART| is a Cartesian (nearest-neighbor)
topology. Special topologies can be defined with \verb|MPI_GRAPH|. In
this paper we only focus on the MPI parallelization of the Monte Carlo
\vamp. A specific profiling of run times of our MPI parallelization
could reveal specific topological structures in the communication
which might offer potential for improvement of speedups. This,
however, is beyond the scope of this paper. 

Messages are passed between sender(s) and receiver(s) inside a
communicator or between communicators where the following
communication modes are available:
\begin{description}
\item[non-blocking] A non-blocking procedure returns directly after
  initiating a communication process. The status of communication must
  be checked by the user. 
\item[blocking] A blocking procedure returns after the communication
  process has completed. 
\item[point-to-point] A point-to-point procedure communicates
  beetween a single receiver and single sender. 
\item[collective] A collective procedure communicates with the
  complete process group. Collective procedures must appear in the
  same order on all processes. 
\end{description}
The standard distinguishes between blocking and non-blocking
point-to-point or collective communications. A conciser program flow
and an increased speedup are advantages of non-blocking over blocking
communication.  

In order to ease the startup of a parallel application, the standard
specifies the startup command \verb|mpiexec|. However, we recommend
the de-facto standard startup command \verb|mpirun| which is in
general part of a MPI-library. In this way, the user does not have to
bother with the quirks of the overall process management and
inter-operability with the operating system, as this is then covered
by \verb|mpirun|. Furthermore, most MPI-libraries support interfaces
to cluster software, e.g. \textsc{SLURM}, \textsc{Torque},
\textsc{HTCondor}.

In summary, we do not use process and (shared-)memory management,
topologies and the advanced error handling of MPI, which we postpone
to a future work. 

\subsubsection{Implementation Details of the MPI Parallelization}

In this subsection, we give a short overview of the technical details
of the implementation to show and explain how the algorithm works in
detail. 

In order to minimize the communication time $T_{\mathrm{c}}$, we only
communicate runtime data which cannot be provided \textit{a priori}
by \wz's pre-integration setup through the interfaces of the
integrators. Furthermore, we expect that the workers are running
identical code up to different communication-based code branches. The
overall worker setup is externally coordinated by the MPI-library
provided process manager \verb|mpirun|. As mentioned in the last
subsection, most MPI libraries have interfaces supporting cluster
software to automatically do the steering between the different nodes,
e.g.~\textsc{SLURM}, \textsc{Torque} or \textsc{HTCondor}.  

In order to enable file input/output (I/O), in particular to allow the
setup of a process, without user intervention, we implement the
well-known master-slave model. The master worker, specified by rank 0,
is allowed to setup the matrix-element library and to provide the
phase-space configuration (or to write the grid files of \vamp) as
those are involved with heavy I/O operations. The other workers
function solely as slave workers supporting only integration and event
generation. Therefore, the slave workers have to wait during the setup
phase of the master worker. We implement this dependence via a
blocking call to \verb|MPI_BCAST| for all slaves while the master is
going through the setup steps.  As soon as the master worker has
finished the setup, the master starts to broadcast a simple logical
which completes the blocked communication call of the slaves allowing
the execution of the program to proceed. The slaves are then allowed
to load the matrix-element library and read the phase-space
configuration file \textit{in parallel}. The slave setup adds a major
contribution to the serial time, mainly out of our control as the
limitation of the parallel setup of the slave workers are imposed by
the underlying filesystem and/or operating system, since all the
workers try to read the files simultaneously. We expect that the
serial time is increased at least by the configuration time of \wz
without building and making the matrix-element library and configuring
the phase-space. Therefore, we expect the configuration time at least
to increase linearly with the number of workers. 

In the following, we outline the reasoning and implementation
details. At the beginning of an iteration pass of \vamp, we broadcast
the current grid setup and the channel weights from the master to all
slave workers. For this purpose, the MPI protocol defines collective
procedures, e.g. \verb|MPI_BCAST| for broadcasting data from one
process to all other processes inside the \verb|communicator|. The
multi-channel formulas Eq.~\eqref{eq:vamp-master} and
Eq.~\eqref{eq:g(p)} force us to communicate each grid\footnote{The
grid type holds information on the binning $x_i$, the number of
dimensions, the integration boundaries and the jacobian.} to every
worker. The details of an efficient communication algorithm and its
implementation is part of the actual MPI implementation (most notably
the \textsc{OpenMPI} and \textsc{MPICH} libraries) and no concern of
us. After we have communicated the grid setup using the collective
procedure \verb|MPI_BCAST|, we let each worker sample over a
predefined set of phase-space channels. Each worker skips its
non-assigned channels and advances the stream of random numbers to the
next substream such as it would have used them for sampling. However,
if we can defer the parallelization to \vegas, we spawn a
\verb|MPI_BARRIER| waiting for all other workers to finish their
computation until the call of the barrier and start with
parallelization of the channel over its grid. When all channels have
been sampled, we collect the results from every channel and combine
them to the overall estimate and variance. We apply a master/slave
chain of communication where each slave sends his results to the
master as shown in Listing~\ref{lst:vamp2-integrate-collect}. For this
purpose the master worker and the slave worker execute different parts
of the code. The computation of the final results of the current pass
is then exclusively done by the master worker. Additionally, the master
writes the results to a \vamp grid file for the case that the
computation is interrupted and should be restarted after the latest
iteration (adding extra serial time to \wz\ runs). 

\begin{lstlisting}[caption={Collecting the results of the multi-channel
    computation on the master worker with rank $0$.},
  language={[08]Fortran},
  label=lst:vamp2-integrate-collect]
subroutine vamp2_integrate_collect ()
  ! ...
  do ch = 1, self%config%n_channel
     if (self%integrator(ch)%is_parallelizable ()) cycle
     worker = map_channel_to_worker (ch, n_size)
     result = self%integrator(ch)%get_result ()
     if (rank == 0) then
        if (worker /= 0) then
           call result%receive (worker, ch)
           call self%integrator(ch)%receive_distribution (worker, ch)
           call self%integrator(ch)%set_result (result)
        end if
     else
        if (rank == worker) then
           call result%send (0, ch)
           call self%integrator(ch)%send_distribution (0, ch)
        end if
     end if
  end do
  ! ...
end subroutine vamp2_integrate_collect
\end{lstlisting} 

In order to employ the \vegas parallelization from~\cite{Kreckel_1997}
we divide the $r$ space into a parallel subspace $r_{\parallel}$ with
dimension $d_{\parallel} = \lfloor d / 2 \rfloor$ over which we
distribute the workers. We define the left-over space $r_{\perp} = r
\setminus r_{\parallel}$ as perpendicular space with dimension
$d_{\perp} = \lceil d / 2 \rceil$. Assigning to each worker a subspace
$r_{\parallel, i} \subset r_{\parallel}$, the worker samples
$r_{\parallel, i} \otimes r_{\perp}$. For the implementation we split
the loop over the cells in $r$-space into an outer parallel loop and
an inner perpendicular loop. In the outer parallel loop the
implementation descends in the inner loop only when worker and
corresponding subspace match, if not, we advance the state of the
\begin{lstlisting}[caption={Iteration over $r$-space and advancing the
    random number generator.},
  language={[08]Fortran}]
  if (self%is_parallelizable ()) then
     do k = 1, rank
        call increment_box_coord (self%box(1:n_dim_par), box_success)
        if (.not. box_success) exit
     end do
     select type (rng)
     type is (rng_stream_t)
        call rng%advance_state (self%config%n_dim * self%config%calls_per_box&
             & * self%config%n_boxes**(self%config%n_dim - n_dim_par) * rank)
     end select
  end if
  loop_over_par_boxes: do while (box_success)
     loop_over_perp_boxes: do while (box_success)
        do k = 1, self%config%calls_per_box
           call self%random_point (rng, x, bin_volume)
           fval = self%jacobian * bin_volume * func%evaluate (x)
           ! ...
        end do
        ! ...
        call increment_box_coord (self%box(n_dim_par + 1:self%config&
             &%n_dim), box_success)
     end do loop_over_perp_boxes
     shift: do k = 1, n_size
        call increment_box_coord (self%box(1:n_dim_par), box_success)
        if (.not. box_success) exit shift
     end do shift
     if (self%is_parallelizable ()) then
        select type (rng)
        type is (rng_stream_t)
           call rng%advance_state (self%config%n_dim * self%config%calls_per_box&
                & * self%config%n_boxes**(self%config%n_dim - n_dim_par) * (n_size - 1))
        end select
     end if
  end do loop_over_par_boxes
\end{lstlisting}
random number generator by the number of sample points in
$r_{\parallel, i} \otimes r_{\perp}$, where $i$ is the skipped outer
loop index.

After sampling over the complete $r$-space the results of the subsets
are collected. All results are collected by reducing them by an
operator, e.g. \verb|MPI_SUM| or \verb|MPI_MAX| with
\verb|MPI_REDUCE| (reduction here is meant as a concept from
functional programming where data reduction is done by reducing a set
of numbers into a smaller set of numbers via a function or an
operator). The application of such a procedure from a MPI 
library is in general more efficient than a self-written procedure. We
implemented all communication calls as non-blocking, i.e. the called
procedure will directly return after setting up the communication.
The communication itself is done in the background, e.g. by an
additional communication thread. The details are provided in the
applied MPI library. To ensure the completion of communication a call
to \verb|MPI_WAIT| has to be done. 

When possible, we let objects directly communicate by
\ffortran\ type-bound procedures, e.g. the main \vegas\ grid object,
\verb|vegas_grid_t| has \verb|vegas_grid_broadcast|. The latter
broadcasts all relevant grid information which is not provided by the
API of the integrator. We have to send the number of bins to all
processes before the actual grid binning happnes, as the size of the
grid array is larger than the actual number of bins requested by
\vegas.\footnote{The size of the grid array is set to a
  pre-defined or user-defined value. If only the implementation
  switches to stratified sampling, the number of bins is adjusted to
  the number of boxes/cells and, hence, does not necessarily match the
  size of the grid array.}  

\begin{lstlisting}[caption={Broadcast grid information using blocking
    and non-blocking
    procedures.},language={[08]Fortran}]
subroutine vegas_grid_broadcast (self)
  class(vegas_grid_t), intent(inout) :: self
  integer :: j, ierror
  type(MPI_Request), dimension(self%n_dim + 4) :: status
  ! Blocking
  call MPI_Bcast (self%n_bins, 1, MPI_INTEGER, 0, MPI_COMM_WORLD)
  ! Non blocking
  call MPI_Ibcast (self%n_dim, 1, MPI_INTEGER, 0, MPI_COMM_WORLD, status(1))
  call MPI_Ibcast (self%x_lower, self%n_dim, &
       & MPI_DOUBLE_PRECISION, 0, MPI_COMM_WORLD, status(2))
  call MPI_Ibcast (self%x_upper, self%n_dim, &
       & MPI_DOUBLE_PRECISION, 0, MPI_COMM_WORLD, status(3))
  call MPI_Ibcast (self%delta_x, self%n_dim, &
       & MPI_DOUBLE_PRECISION, 0, MPI_COMM_WORLD, status(4))
  ndim: do j = 1, self%n_dim
     call MPI_Ibcast (self%xi(1:self%n_bins + 1, j), self%n_bins + 1,&
          & MPI_DOUBLE_PRECISION, 0, MPI_COMM_WORLD, status(4 + j))
  end do ndim
  call MPI_Waitall (self%n_dim + 4, status, MPI_STATUSES_IGNORE)
end subroutine vegas_grid_broadcast
\end{lstlisting} 

Further important explicit implementations are the type-bound
procedures \verb|vegas_send_distribution|/
\verb|vegas_receive_distribution| and
\verb|vegas_result_send|/\verb|vegas_result_receive| which are needed
for the communication steps involved in \vamp in order to keep  
the \vegas integrator objects encapsulated (i.e. preserve their
\verb|private| attribute).

Beyond the inclusion of non-blocking collective communication we
choose as a minimum prerequisite the major version 3 of MPI for better
interoperability with \fortran\ and its conformity to the \ffortran 
\texttt{+ TS19113} (and later) standard~\cite{MPI_3.1},
Sec. 17.1.6. This, e.g., allows for MPI-derived type comparison as
well as asynchronous support (for I/O or non-blocking communication). 

A final note on the motivation for the usage of non-blocking procedures. 
Classic (i.e. serial) Monte Carlo integration exhibits no need for
in-sampling communication in contrast to classic application of
parallelization, e.g.~solving partial differential equations. For the
time being, we still use non-blocking procedures in \vegas for future
optimization, but in a more or less blocking fashion, as most
non-blocking procedures are followed by a \verb|MPI_WAIT|
procedure. However, the multi-channel ansatz adds sufficient
complexity, as each channel itself is an independent Monte Carlo
integration. A typical use case is the collecting of already sampled
channels while still sampling the remaining channels as it involves
the largest data transfers in the parallelization setup. Here, we
could benefit most from non-blocking communication. To implement these
procedures as non-blocking necessitates a further refactoring of the
present multi-channel integration of \wz, because in that case the
master worker must not perform any kind of calculation but should only
coordinate communication. A further constraint to demonstrate the
impact of turning many of our still blocking communication into a
non-blocking one is the fact that at the current moment, there do not
exist any profilers compliant with the MPI3.1 status that support
\ffortran. Therefore, we have to postpone the opportunity to show
the possibility of completely non-blocking communication in our
setup.

\subsubsection{Speedup and Results}
\label{sec:speedup}

\begin{figure}
  \centering
  \includegraphics[width=\textwidth]{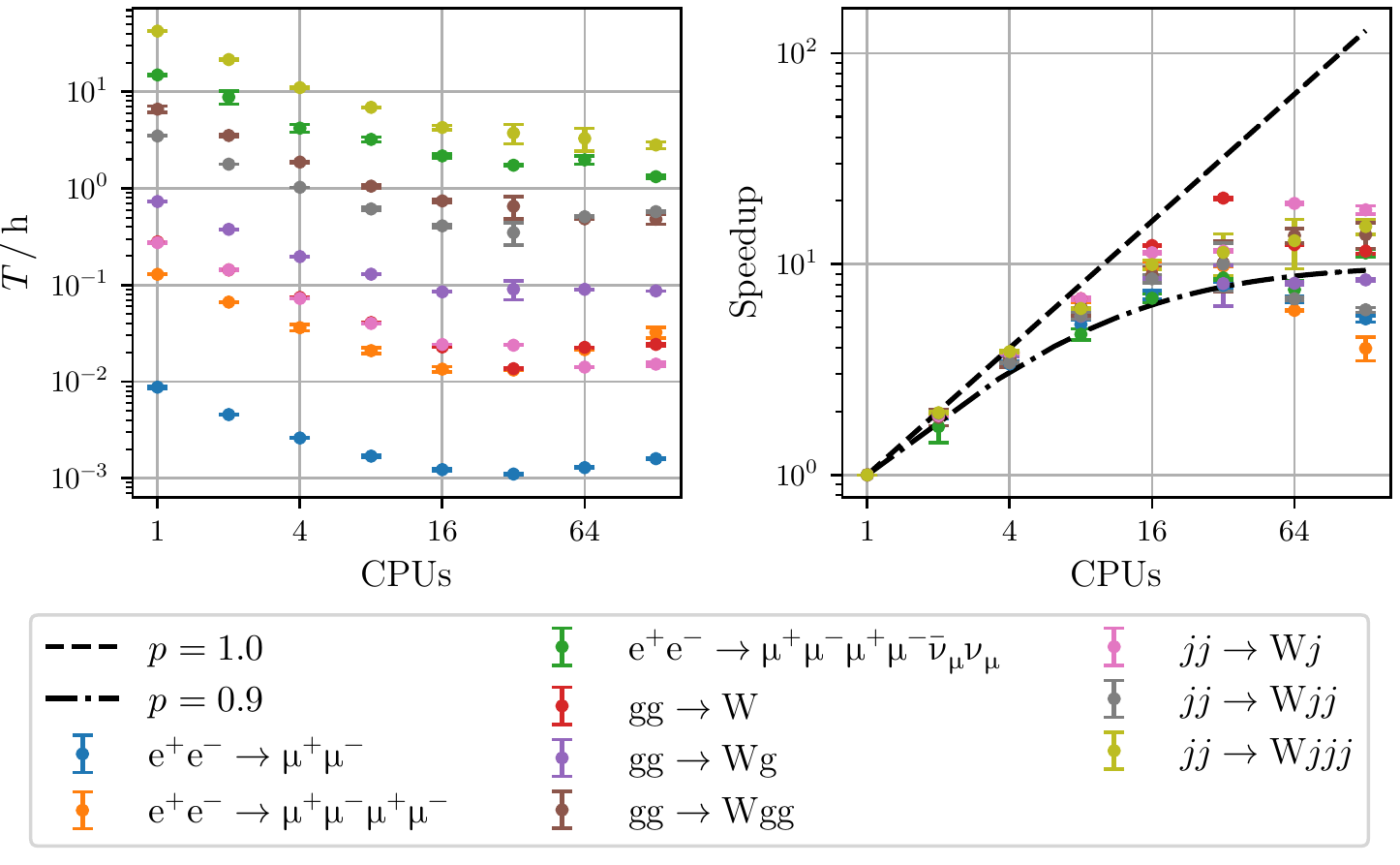} 
  \caption{We show the overall computation time for different complex
    processes for different numbers of participating CPUs (left
    panel). The numbers of CPUs is chosen as a power of 2. In the
    right panel, we plot the speedup of the processes and compare them
    to ideal of Amdahl's Law with parallelizable fractions of 1.0
    (dashed) and 0.9 (dash-dotted).}
  \label{fig:benchmark}
\end{figure}

In order to assess the efficiency of our parallelization, we compare
the two modes, the traditional serial \vamp\ implementation and our
new parallelized implementation. In the following, we study different
processes at different levels of complexity in order to investigate
the scaling behavior of our parallel integration algorithm. The
process $\Pep\Pem \to \APmuon\Pmuon$ at energies below the $Z$ resonance
has only one phase-space channel ($s$-channel photon exchange and its
integration is adapted over one grid. Parallelization is done in
\vegas over stratification space. The process $\Pep\Pem \to \APmuon\Pmuon
\APmuon\Pmuon\Pnum\APnum$ with its complicated vector-boson
interactions gives rise to $\mathcal{O}(3000)$ phase-space channels.  
With overall  $\mathcal{O}(10^{6})$ number of calls for the process,
each phase-space channel is sampled (in average) by
$\mathcal{O}(10^2)$ calls suppressing the stratification space of all
grids. Therefore, parallelization is done over the more coarse
phase-space channel loop. The last process we investigate, $\Pep\Pem
\to \APmuon\Pmuon\APmuon\Pmuon$, has $\mathcal{O}(100)$ phase-space
channels where we expect for some grids a distinct stratification space
(at least two cells per dimension) allowing \wz\ to switch between
\vegas and multi-channel parallelization. All but the first trivial
examples are taken from~\cite{Kilian:2007gr} mimicking real-world
application. The results are shown in Fig.~\ref{fig:benchmark}.

\begin{figure}
  \centering
  \includegraphics[width=\textwidth]{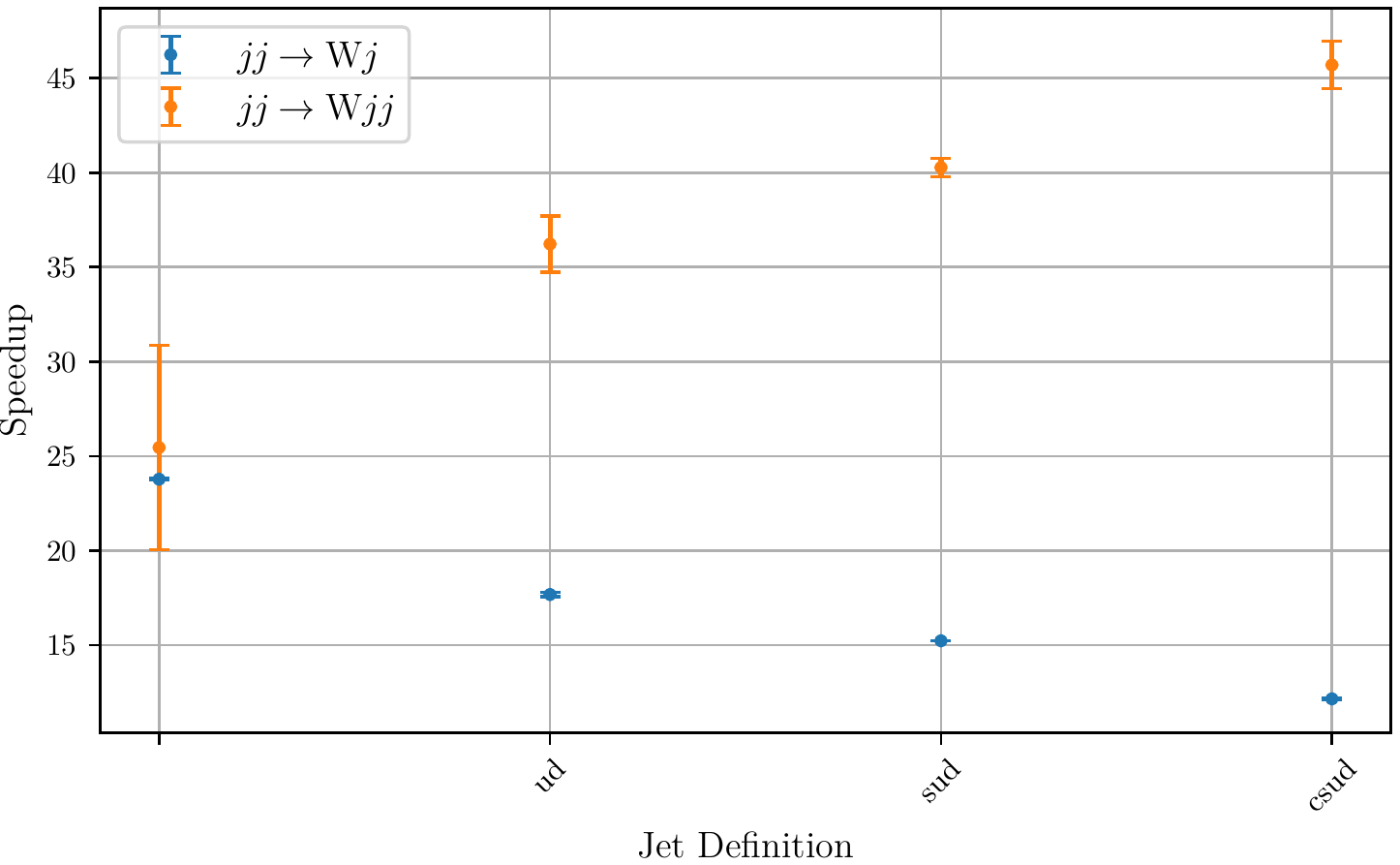}
  \caption{We show the speedup for a jet process with increasing
    flavor content and a fixed numbers of CPUs of 60.} 
  \label{fig:flavor}
\end{figure}

Furthermore, we are interested in the behavior for increasing
complexity of a single process, e.g. increasing (light) flavor content
of processes with multiple jets. For the two processes, $jj \to
\PWm (\to \Pem \bar{\nu}_e) + \{j, jj\}$ we increase the number of
massless quark flavors in the content of the jets. The results in
Fig.~\ref{fig:flavor} indicate that for a single final-state jet more
flavor content (and hence more complicated matrix elements) lead to
lower speedups. For two (and more) final state jets the speedups
increases with the multiplicity of light quarks in the jet definition.
This means that possibly for smaller matrix elements there is a
communication overhead when increasing the complexity of the matrix
element, while for the higher multiplicity process and many more
phase-space channels, improvement in speedup can be achieved.

\begin{figure}
  \centering
  \includegraphics[width=\textwidth]{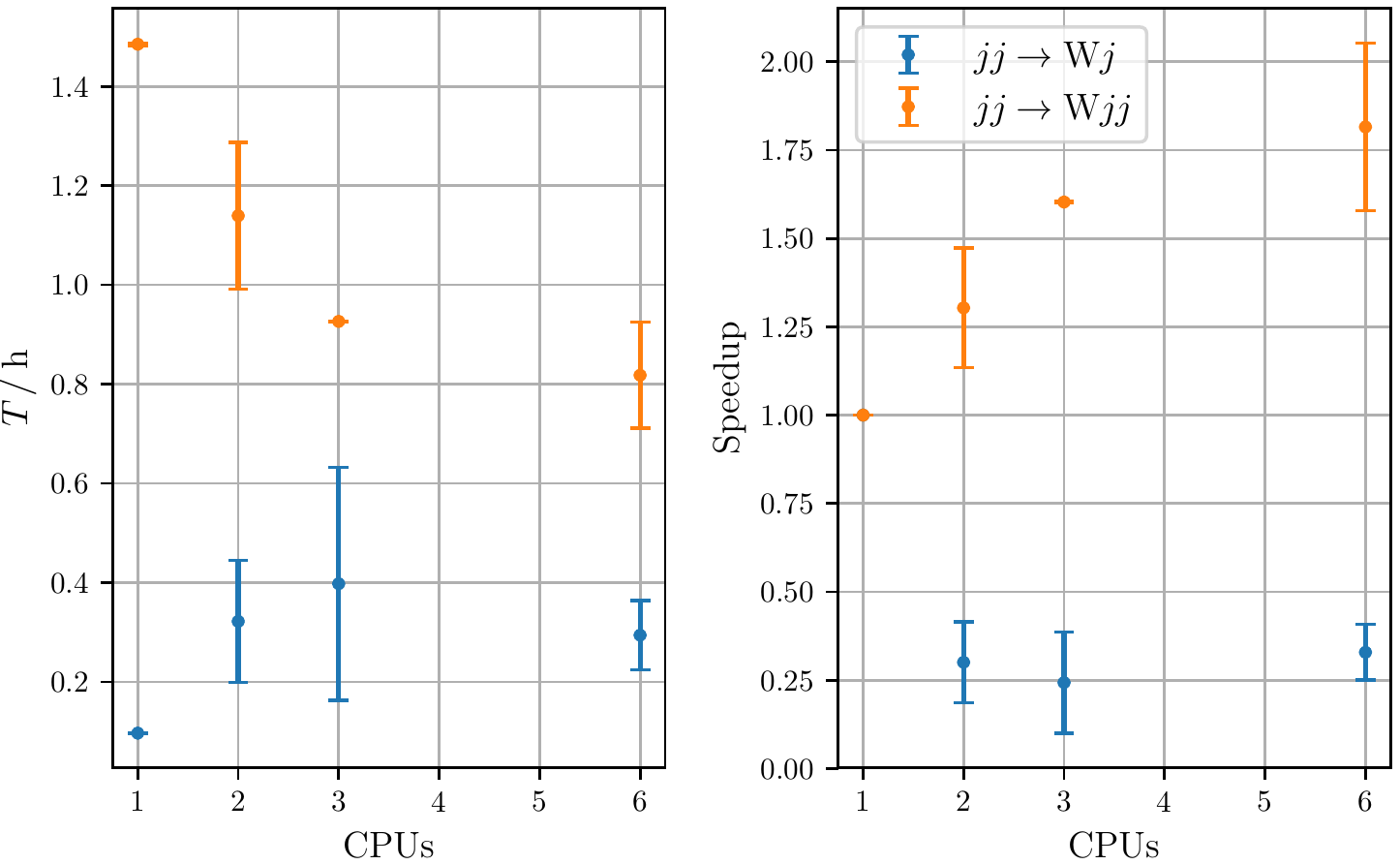}
  \caption{These panels show the speedup for a fixed number of workers
    with increasing number of threads.} 
  \label{fig:var-hybrid}
\end{figure}

\begin{figure}
  \centering
  \includegraphics[width=\textwidth]{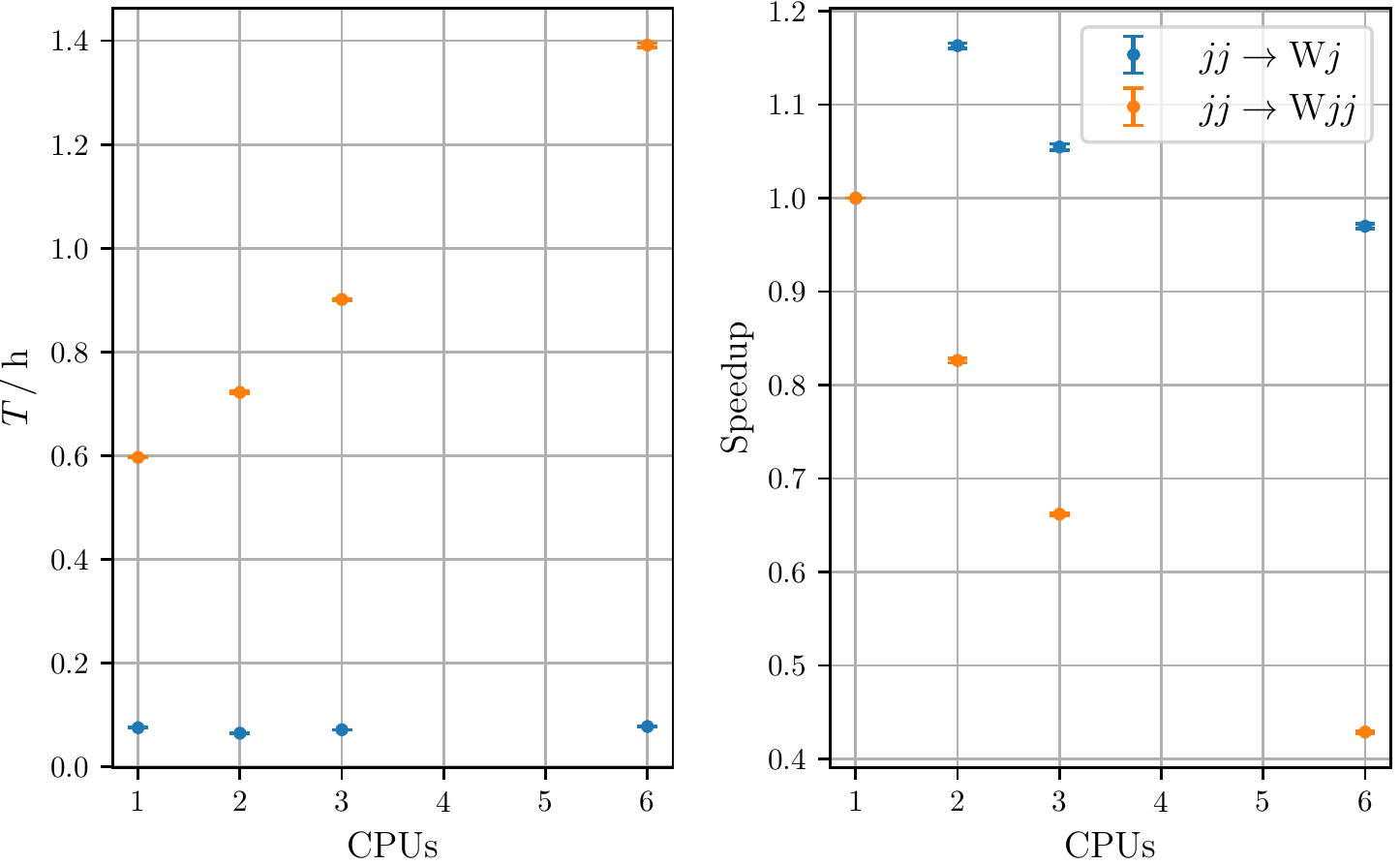}
  \caption{We show the speedup at an overall fixed numbers of CPUs
    (60) involved in the parallelization. We distribute the CPUs among
    MPI and OpenMP parallelization. For the latter we have to respect
    the node structure where each node consists of two CPUs each with
    6 cores handling up to 6 threads without any performance penalty.} 
  \label{fig:fixed-hybrid}
\end{figure}

We benchmark the processes on the high performance cluster of the
University of Siegen (Hochleistungrechner Universität
Siegen, HorUS) which provides 34 Dell PowerEdge C6100 each containing
4 computing nodes with 2 CPUs. The nodes are connected by gigabit
ethernet and the CPUs are Intel Xeon X5650 with 6 cores each
\SI{2.66}{\giga\hertz} and \SI{128}{\mebi\byte} cache. We employ two
different \wz builds, a first one only with \textsc{OpenMPI 2.1.2},
and a second one with additional OpenMP support testing the
hybrid parallelization. The HorUS cluster utilizes \textsc{SLURM
17.02.2} as batch and allocation system allowing for easy source
distribution. We run \wz using the MPI-provided run manager
\verb|mpirun| and measure the run-time with the GNU time command tool,
and we average over three independent runs for the final result.  We
measure the overall computation time of a \wz run including the
complete process setup with matrix-element generation and phase-space
configuration. It is expected that the setup step gives rise to the
major part of the serial computation of \wz, and also the I/O
operations of the multi-channel integrator, which saves the grids
after each integration iteration.  As this is a quasi-standard, we
benchmark over $N_{\text{CPU}}$ in powers of \num{2}. Given the
architecture of the HorUS cluster with its double hex-cores,
benchmarking in powers of \num{6} would maybe be more appropriate for
the MPI-only measurements. We apply a node-specific setup for the
measurement of the hybrid parallelization. Each CPU can handle up to
six threads without any substantial throttling. We operate over
$N_{\text{Threads}} = \{1, 2, 3, 6\}$ with either fixed overall number
of involved cores, $N_{\text{Worker}} = \{60, 30, 20, 10\}$, with
results shown in Fig.~\ref{fig:var-hybrid}, or with fixed number of
workers $N_{\text{Worker}} = 20$, with results shown in
Fig.~\ref{fig:fixed-hybrid}.

Coming back to Fig.~\ref{fig:benchmark} showing the results of the
benchmark measurement for MPI: The process $\Pep \Pem \to
\APmuon\Pmuon$ saturates for $N > 32$. The serial
runtime of \wz is dominating for that process with its two-dimensional
integration measure (without beam structure functions) where the MC
integration is anyways inferior to classical integration techniques. 
The process $\Pep\Pem \to \APmuon\Pmuon\APmuon\Pmuon$ showing mixed
multi-channel and \vegas parallelization, however, also saturates for
$N > 32$. The multi-channel parallelizable process $\Pep\Pem \to
\APmuon\Pmuon\APmuon\Pmuon\Pnum\APnum$
achieves a higher speedup but with decreasing slope. The overall
speedup plot indicates a saturation beginning roughly at $N > 32$
where serial time and communication start to dominate. We conclude
that \wz embarks a parallelization fraction higher than at least
\SI{90}{\percent} for MPI. In the appendix \ref{sec:results} we
present tables that show the actual physical runtimes for the
different processes under consideration.

\subsection{Alternative algorithm for phase-space generation}
\label{sec:wood2}

Profiling of the code reveals that for the moment the main bottleneck
that inhibits speedups beyond $n=100$ is the initial construction of
the phase-space configurations, i.e. the phase-space channels and
their parameterizations (the determination of the best mappings to be
done for each channel or class of channels) which \wz\ constructs from
forests of Feynman tree-graphs.  Resembling the language of
Ref.~\cite{Boos:1999qc}, this construction algorithm is called 
\emph{wood}. This \emph{wood} algorithm takes into account the model
structure, namely the three-point vertices to find resonant
propagators, the actual mass values (to find collinear and soft
singularities and to map mass edges), and the process energy.  It
turns out that while the default algorithm used in \wz\ yields a good
sample of phase-space channels to enable adaptive optimization, it has
originally not been programmed in an efficient way. Though it is a
recursive algorithm, it does not work on directed acyclical graphs
(DAGs) like \om\ to avoid all possible redundancies.

Therefore, a new algorithm, \emph{wood2}, has been designed in order
to overcome this problem.  Instead of constructing the forest of
parameterizations from the model, it makes use of the fact that the
matrix elements constructed optimally by \om\ in form of a DAG already
contain all the necessary information with the exception
of the numerical values for masses and collider energy.  Thus, instead
of building up the forest again, the algorithm takes a suitable
description of the set of trees from \om\ and applies the elimination
and simplification algorithm, in order to yield only the most relevant
trees as phase-space channels. As it turns out, even in a purely
serial mode, the new implementation performs much better and thus
eliminates the most important source of saturation in speedup. Another
benefit of the new algorithm is that it is much less memory-hungry
than the original one which could have become a bottleneck for the
traditional algorithm for complicated processes in very complicated
models (e.g. extensions of the MSSM).

%%%%%

\section{Conclusions and Outlook}
\label{sec:conclusions}

Monte-Carlo simulations of elementary processes are an essential
prerequisite for successful physics analyses at present and future
colliders.  High-multiplicity processes and high precision in signal
and background detection put
increasing demands on the required computing resources.  One
particular bottleneck is the multi-dimensional phase-space integration
and the task of automatically determining a most efficient sampling
for (unweighted) event generation.  In this paper, we have described
an efficient algorithm that employs automatic iterative adaptation in
conjunction with parallel evaluation of the numerical integration and
event generation.

The parallel evaluation is based on the paradigm of the Message
Passing Interface Protocol (MPI) in conjunction with OpenMP
multi-threading. For the concrete realization, the algorithm has been
implemented within the framework of the multi-purpose event
generator \wz.  The parallelization support for MPI or OpenMP can be
selected during the configure step of \wz.  The new code constitutes a
replacement module for the \vamp\ adaptive multi-channel integrator
which makes active use of modern features in the current MPI-3.1
standard.  Our initial tests for a variety of benchmark physics
processes demonstrate a speedup by a factor $>10$
with respect to serial evaluation.  The best results have been
achieved by MPI parallelization.  The new implementation has been
incorporated in the release version \wz~\textsc{v2.6.4}.

We were able to show that, in general, hybrid parallelization with
\textsc{OpenMP} and \textsc{MPI} leads to a speedup which is
comparable to \textsc{MPI} parallelization alone.  However, combining
both approaches is beneficial for tackling
memory-intense processes, such as 8- or 10-particle processes.
Depending on a particular computing-cluster topology, the latter
approach can allow for a more efficient use of the memory locally available
at a computing node.  In the hybrid approach,
\wz\ is parallelized on individual multi-core nodes via
\textsc{OpenMP} multithreading, while distinct computing nodes
communicate with each other via \textsc{MPI}.  The setup of the system 
allows for sufficient flexibility to make optimal use of both
approaches for a specific problem.

The initial tests point to further possibilities for
improvement, which we foresee for future development and refinements
of the implementation.  A server/client structure should give the
freedom to re-allocate and assign workers dynamically during a
computing task, and thus make a more efficient use of the available
resources.  Further speedup can be expected from removing various
remaining blocking communications and replacing them by non-blocking
communication while preserving the integrity of the calculation.
Finally, we note that the algorithm shows its potential for
calculations that spend a lot of time in matrix-element evaluation.
For instance, in some tests of NLO QCD processes we found that the
time required for integration could be reduced from the order of a
week down to a few hours.  We defer a detailed benchmarking of
such NLO processes to a future publication.

%%%%%

\section{Acknowledgements}
The authors want to thank Bijan Chokouf\'{e} Nejad, Stefan Hoeche and
Thorsten Ohl for helpful and interesting discussions. For their
contributions to the new phase-space construction algorithm (known as
\emph{wood2}) we give special credits to Manuel Utsch and Thorsten
Ohl.

\vspace{5mm}

\appendix

\section{Results}
\label{sec:results}

The measured duration of the benchmarks are listed in the
tables~\ref{tab:leptonic},~\ref{tab:hadronic-jet},~\ref{tab:hadronic-quark-gluon}
and~\ref{tab:hadronic-jet-flavor}. It can be seen that the processes
with a longer duration are subject to a large run-time fluctuation. We
were able to reproduce this behavior with a variety of processes and
attribute it back to the messy benchmark environment. The HoRUS of the
University of Siegen is a university-wide cluster and is usually well
utilized by other research groups, too. Therefore, a clean and
reproducible measurement of longer runtime was difficult to make
because on average we do not have the necessary computing power or
network performance for us alone, which is reflected in particular in
longer runtime. 

\begin{table}
  \centering
  \sisetup{
  scientific-notation=false,
  separate-uncertainty=true,
}
\begin{tabular}{
  S
  S %[table-format=2.3e+1]
  S %[table-format=3.4e+1]
  S %[table-format=5.6e+1]
  }
  \toprule
  {$N_{\text{CPU}}$} & {$T(\Pelectron \APelectron \to \Pmuon \APmuon) \SI{}{\per\second}$} & {$T(\Pelectron \APelectron \to \Pmuon \APmuon \Pmuon \APmuon) \SI{}{\per\second}$} & {$T(\Pelectron \APelectron \to \Pmuon \APmuon \Pmuon \APmuon \Pnum \APnum) \SI{}{\per\second}$} \\
  \midrule
 1   & 31.6(11) & 465.0(29)  & 53813.8(9204)  \\
 2   & 16.4(1)  & 239.8(4)   & 31768.7(50680) \\
 4   & 9.4(1)   & 130.3(99)  & 15145.2(13699) \\
 8   & 6.1(2)   & 75.3(52)   & 11579.7(6955)  \\
 16  & 4.4(1)   & 48.5(33)   & 7815.4(3466)  \\
 32  & 4.0(1)   & 47.5(4)    & 6257.2(874)   \\
 64  & 4.6(1)   & 77.1(3)    & 7110.4(6931)  \\
 128 & 5.7(1)   & 116.7(151) & 4763.8(1907)  \\
  \bottomrule
\end{tabular}
  \caption{Computation time of the leptonic processes.}
  \label{tab:leptonic}
\end{table}

\begin{table}
  \centering
  \sisetup{
  scientific-notation=false,
  separate-uncertainty=true,
}
\begin{tabular}{
  S
  S %[table-format=2.3e+1]
  S %[table-format=3.4e+1]
  S %[table-format=5.6e+1]
  }
  \toprule
  {$N_{\text{CPU}}$} & {$T(\Pjet \Pjet \to \PW \Pjet) \SI{}{\per\second}$} & {$T(\Pjet \Pjet \to \PW \Pjet \Pjet) \SI{}{\per\second}$} & {$T(\Pjet \Pjet \to \PW \Pjet \Pjet \Pjet) \SI{}{\per\second}$} \\
  \midrule
  1   & 987.8(94) & 12561.4(776) & 153072.1(14984) \\
  2   & 517.2(47) & 6413.5(312)  & 77640.6(6869)   \\
  4   & 261.5(12) & 3699.7(15)   & 39812.5(3898)   \\
  8   & 144.2(19) & 2208.6(834)  & 24867.7(1960)   \\
  16  & 87.3(01)  & 1475.3(650)  & 15390.6(7696)   \\
  32  & 85.7(07)  & 1258.0(3218) & 13475.7(30259)  \\
  64  & 50.9(01)  & 1834.3(304)  & 11854.4(31335)  \\
  128 & 54.6(23)  & 2069.1(519)  & 10152.1(8135)   \\
  \bottomrule
\end{tabular}
  \caption{Computation time of $\Pjet \Pjet \to \PWm(\to \Pelectron
    \APnue) + n\Pjet$ processes.} 
  \label{tab:hadronic-jet}
\end{table}

\begin{table}
  \centering
  \sisetup{
  scientific-notation=false,
  separate-uncertainty=true,
}
\begin{tabular}{
  S
  S %[table-format=2.3e+1]
  S %[table-format=3.4e+1]
  S %[table-format=5.6e+1]
  }
  \toprule
  {$N_{\text{CPU}}$} & {$T(\Pgluon \Pgluon \to \PW \Pquark \APquark) \SI{}{\per\second}$} & {$T(\Pgluon \Pgluon \to \PW \Pquark \APquark \Pgluon) \SI{}{\per\second}$} & {$T(\Pgluon \Pgluon \to \PW \Pquark \APquark \Pgluon \Pgluon) \SI{}{\per\second}$} \\
  \midrule
  1   & 1008.0(16) & 2634.2(300) & 23823.2(18765) \\
  2   & 516.4(14)  & 1357.4(05)  & 12712.5(4251)  \\
  4   & 267.8(39)  & 709.1(96)   & 6745.1(1375)   \\
  8   & 147.0(08)  & 466.1(10)   & 3800.5(1743)   \\
  16  & 82.3(03)   & 305.8(25)   & 2673.5(835)    \\
  32  & 49.0(07)   & 326.0(712)  & 2352.7(6030)   \\
  64  & 81.6(06)   & 324.9(26)   & 1743.3(278)    \\
  128 & 87.5(25)   & 313.0(03)   & 1727.5(1949)   \\
  \bottomrule
\end{tabular}
  \caption{Computation time of $\Pgluon \Pgluon \to \PWm(\to
    \Pelectron \APnue)\Pquark \APquark + n\Pgluon$ processes.} 
  \label{tab:hadronic-quark-gluon}
\end{table}

\begin{table}
  \centering
  % N d ud sud csud
\sisetup{
  tight-spacing=true,
  separate-uncertainty=true,
}
\begin{tabular}{
  l
  S
  S
  S
  S
  S
  }
  \toprule
  process & {$N_{\text{CPU}}$} & {$T(\lbrace \Pdown \rbrace) \SI{}{\per\second}$} & {$T(\lbrace \Pup, \Pdown \rbrace) \SI{}{\per\second}$} & {$T(\lbrace \Pup, \Pdown, \Pstrange \rbrace) \SI{}{\per\second}$} &  {$T(\lbrace \Pup, \Pdown, \Pstrange, \Pcharm \rbrace) \SI{}{\per\second}$} \\
  \midrule
  % jjWj
  \multirow{2}{*}{$\Pjet \Pjet \to \PW \Pjet$} & 1 & 1589.1(30) & 2189.3(83) & 2730.9(12) & 3440.6(112) \\
          & 60 & 66.8(0) & 123.8(7) & 179.3(1) & 283.1(8) \\
          % jjWjj
  \midrule
  \multirow{2}{*}{$\Pjet \Pjet \to \PW \Pjet \Pjet$} & 1 & 11075.0(23477) & 29405.0(12069) & 49151.5(5856) & 91677.3(25027) \\
          & 60 & 435.1(3) & 811.7(4) & 1220.4(8) & 2006.0(5) \\
  \bottomrule
\end{tabular}
  \caption{Computation time over increasing flavor content.
    The upper two lines are for the process $\Pproton \Pproton \to \PW
    \Pjet$, the lower two for the process $\Pproton \Pproton \to \PW
    \Pjet \Pjet$, respectively. The second column gives the number of
    CPU cores, the following columns are the results for an increasing
    number of massless quark flavors in the initial state and jets,
    growing from one ($\Pdown$) to four ($\Pdown, \Pup, \Pstrange,
    \Pcharm$).} 
  \label{tab:hadronic-jet-flavor}
\end{table}

\printbibliography{}
\end{document}